\documentclass[10pt,journal,compsoc]{IEEEtran}
\usepackage{graphicx}
\usepackage{tikz}
\usepackage{comment}
\usepackage{amsmath,amssymb}

\usepackage{hyperref}
\usepackage{color}

\usepackage[accsupp]{axessibility}
\usepackage{multirow}
\usepackage{makecell}

\ifCLASSOPTIONcompsoc
  \usepackage[nocompress]{cite}
\else
  \usepackage{cite}
\fi
\ifCLASSINFOpdf
\else
\fi
\begin{document}

\newcommand{\etal}{\textit{et al.}}

\newsavebox\CBox
\def\textBF#1{\sbox\CBox{#1}\resizebox{\wd\CBox}{\ht\CBox}{\textbf{#1}}}

\title{Learning Spatiotemporal Frequency-Transformer for Low-Quality Video Super-Resolution}

\author{Zhongwei~Qiu,
        Huan~Yang,
        Jianlong~Fu,
        Daochang~Liu,
        Chang~Xu,
        and~Dongmei~Fu% <-this % stops a space
\IEEEcompsocitemizethanks{\IEEEcompsocthanksitem Z. Qiu and D. Fu are with University of Science and Technology Beijing, Beijing 10083, P.R. China. Z. Qiu is also with University of Sydney, Darlington, NSW 2008, Australia.
\protect\\
E-mail: qiuzhongwei@xs.ustb.edu.cn, fdm\_ustb@ustb.edu.cn

\IEEEcompsocthanksitem H. Yang and J. Fu are with Microsoft Research, Beijing 10080, P.R. China. \protect\\
E-mail: huayan@microsoft.com, jianf@microsoft.com

\IEEEcompsocthanksitem D. Liu and C. Xu are with University of Sydney, Darlington, NSW 2008, Australia. \protect\\
E-mail: daochang.liu@sydney.edu.au, c.xu@sydney.edu.au

\IEEEcompsocthanksitem Corresponding author: Huan Yang and Dongmei Fu
}% <-this % stops an unwanted space
% \thanks{Manuscript received April 19, 2022; revised August 26, 2015.}
}
% \thanks{Manuscript received April 19, 2022; revised August 26, 2015.}}


\markboth{Journal of \LaTeX\ Class Files,~Vol.~14, No.~8, August~2015}%
{Shell \MakeLowercase{\textit{et al.}}: Bare Demo of IEEEtran.cls for Computer Society Journals}



\IEEEtitleabstractindextext{%
\begin{abstract}
Video Super-Resolution (VSR) aims to restore high-resolution (HR) videos from low-resolution (LR) videos. Existing VSR techniques usually recover HR frames by extracting pertinent textures from nearby frames with known degradation processes. Despite significant progress, grand challenges are remained to effectively extract and transmit high-quality textures from high-degraded low-quality sequences, such as blur, additive noises, and compression artifacts. In this work, a novel Frequency-Transformer (FTVSR) is proposed for handling low-quality videos that carry out self-attention in a combined space-time-frequency domain. First, video frames are split into patches and each patch is transformed into spectral maps in which each channel represents a frequency band. It permits a fine-grained self-attention on each frequency band, so that real visual texture can be distinguished from artifacts. Second, a novel dual frequency attention (DFA) mechanism is proposed to capture the global frequency relations and local frequency relations, which can handle different complicated degradation processes in real-world scenarios. Third, we explore different self-attention schemes for video processing in the frequency domain and discover that a ``divided attention'' which conducts a joint space-frequency attention before applying temporal-frequency attention, leads to the best video enhancement quality. Extensive experiments on three widely-used VSR datasets show that FTVSR outperforms state-of-the-art methods on different low-quality videos with clear visual margins.
Code and pre-trained models are available at \href{https://github.com/researchmm/FTVSR}{https://github.com/researchmm/FTVSR}.
\end{abstract}

}

\maketitle

\IEEEdisplaynontitleabstractindextext

% \fi
\IEEEpeerreviewmaketitle

\IEEEraisesectionheading{\section{Introduction}
\label{sec:introduction}}

\IEEEPARstart{V}{ideo} super-resolution (VSR) is a fundamental task in computer vision, which aims to restore a sequence of high-resolution (HR) frames from its low-resolution (LR) counterparts. VSR can benefit a broad range of downstream applications, such as high-definition television~\cite{goto2014super} and video surveillance~\cite{zhang2010super}. 
Existing VSR methods focus on leveraging temporal information from sliding windows~\cite{kim20183dsrnet,li2019fast,tian2020tdan,wang2019edvr} or recurrent structures~\cite{chan2021basicvsr,sajjadi2018frame,yi2021omniscient}, and have achieved great success in limited scenarios that usually take LR video frames with known degradation process as inputs.

\begin{figure*}[!t]
\centering
\includegraphics[width=0.95\textwidth]{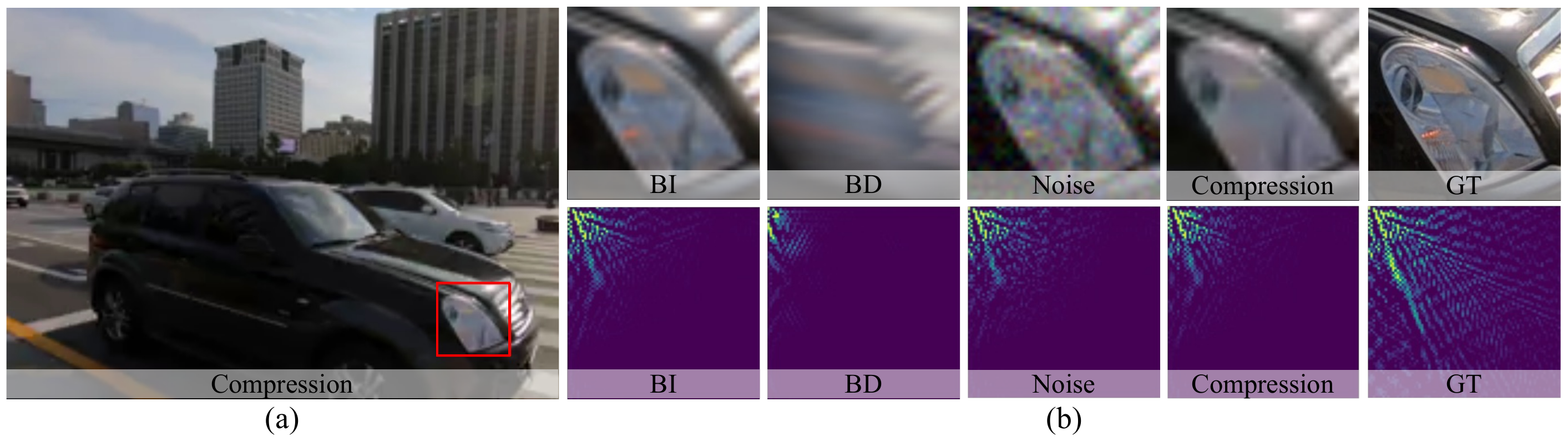}
\vspace{-0.2cm}
\caption{Comparison of low-resolution frames and their spectral maps with different degradations (BI: Bicubic downsampling, BD: Downsampling with blur kernel, Noise: Bicubic downsampling with noise, Compression: Bicubic downsampling with video compression, GT: Ground-truth of high-resolution). (a) The low-resolution frame generated by Bicubic downsampling with video compression. (b) The cropped patches and spectral maps from low-resolution frames.}
\label{fig:compar_lr_videos}
\end{figure*}

However, in real-world scenarios, most videos on the internet or on user devices are stored and transmitted in low-quality formats with unknown degradations. For example, low-resolution videos can be generated with various corruptions (downsampling with blur kernels, noises), and different compression algorithms. These complicated degradations result in different information loss, which makes it more challenging to recover the HR frames from low-quality LR counterparts. As shown in Figure \ref{fig:compar_lr_videos}, the low-quality frames with different degradations show different artifacts. 
Such variational characteristics of low-quality videos make it harder to recover HR frames by using current VSR methods~\cite{chan2021basicvsr,chan2022basicvsr++,liu2022learning} since they are designed with known degradation process.

To handle the problem of unknown degradation, the blind VSR methods~\cite{liu2013bayesian,lee2021dynavsr,pan2021deep} firstly estimate the degradation kernels and noise parameters, then design the special modules with the estimated degradation parameters to remove blur and noise.
However, these designs are specific and not suitable for other degradations in real-world scenes, such as video compression. 
The compression artifacts are generated from the quantification process acting on the image patches transformed by Discrete Cosine Transformation (DCT). Different from the global degradations in blur or noise, compression has different information lost on local patches and can not estimate its degradation process.

\begin{figure*}[!t]
\centering
\includegraphics[width=0.95\textwidth]{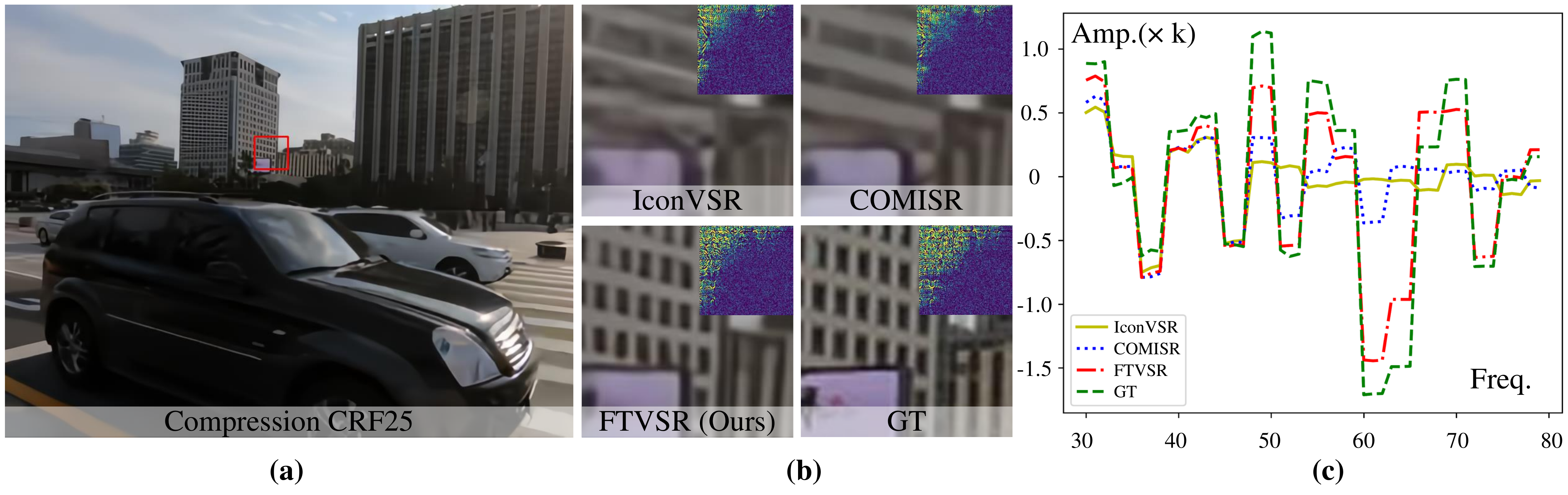}
\vspace{-0.2cm}
\caption{The comparison between FTVSR, IconVSR~\cite{chan2021basicvsr}, and COMISR~\cite{li2021comisr} on compressed LR videos with compression rate of CRF25. (a) The $\times 4$ VSR results generated by FTVSR. (b) Visualizations of zoom-in patches and their DCT-based spectral maps (shown in the top-right corner). FTVSR recovers more high-frequency information than COMISR and IconVSR. (c) The Amplitude-Frequency curves on clipped frequency bands of 30 to 80. FTVSR is superior than other methods to approximate the curve of ground truth.}
\label{fig:teaser}
\end{figure*}

Video compression is one of the commonest sources of degradation in real-world scenarios. For example, the most widely-used video codec H.264 takes a constant rate factor (CRF) varied from $0$ to $51$ as its parameter to control the compression rate and results in varying degrees of degradations. As shown in Figure~\ref{fig:teaser}, the existing SOTA IconVSR method~\cite{chan2021basicvsr} fails to recover pleasant visual results on such compressed LR videos.
The compression artifacts are magnified during the restoration processes since this model takes unseen compression artifacts as common textures. 
Li~\etal~\cite{li2021comisr} retrain the model with compressed videos but achieve limited gain since it is hard for the model to recover high-frequency textures while simultaneously removing the high-frequency compression artifacts. To strengthen the awareness of compression, COMISR~\cite{li2021comisr} predicts detail-aware flow to align features and restore HR frames by a Laplacian enhancement module. However, as shown in Figure~\ref{fig:teaser}, large gaps between the generated frame and the ground truth remain. Thus, compression degradation is difficult to process and needs special designs.

Other real-world VSR methods propose new real-world datasets or pre-processing modules to deal with these problems. RealVSR~\cite{yang2021real} and RealBasicVSR~\cite{chan2022investigating} propose a new dataset for evaluation, while the real HR frames cannot be obtained for training. RealBasicVSR~\cite{chan2022investigating} uses a "Pre-cleaning" module to remove artifacts before applying BasicVSR~\cite{chan2021basicvsr}. However, the "Pre-cleaning" module still cannot solve the video compression problem and has limited generalization ability in real-world scenarios since it needs degradation-cleaning image pairs for training.

To handle these complicated degradations including compression in real-world scenarios, we propose a novel \textbf{F}requency \textbf{T}ransformer for low-quality \textbf{V}ideo \textbf{S}uper-\textbf{R}esolution (\textbf{FTVSR}). 
The insightful idea is to use the Discrete Cosine Transform (DCT) to convert low-quality video frames into a series of frequency-based representations and then design frequency-based attention to enable interactions between frequency features across various frequency bands.
This design brings two key merits: 
1) The patch-based frequency representation treats each frequency band equitably, effectively preserving high-frequency visual information;
2) frequency-based attention guides the generation of high-frequency textures from low-frequency information (e.g., object structure), which can significantly reduce the effect of corruption artifacts.
In order to remove the artifact caused by different degradations, we propose Dual Frequency Attention (DFA), which can capture the local and global frequency relations to better deal with different degradations, such as compression, blur, noise, and so on. 
For example, compression artifacts show different patterns in the local areas of the image and need stronger local attention, while blur-based degradations show similar patterns in the global area and need more global attention.
Additionally, several frequency-based attentions that effectively combine with space and time attention are investigated to further extract the frequency interactions in both the spatial and temporal dimensions for VSR.
Our main contributions can be summarized as follows:
\begin{itemize}
    \item We propose and successfully process VSR in the frequency domain, which is the first VSR work to introduce learning frequency dependencies.
    \item We propose frequency attention to build a frequency Transformer for VSR. Besides, we further propose enhanced dual frequency attention to handle the complicated degradations in the real-world VSR.
    \item We discuss different self-attention strategies for processing frequency-domain video and find that the highest video enhancement quality comes from "divided attention," which first performs a joint space-frequency attention before applying temporal-frequency attention to each frequency band.
    \item Extensive experiments on several widely-used VSR benchmarks with multiple complicated degradations and real-world VSR benchmarks show the superior performance of FTVSR. 
\end{itemize}

Our previous work~\cite{qiu2022learning} in conference proposed FTVSR for compressed video super-resolution, which restores the compressed LR videos by a frequency Transformer. In this work, we tackle the more complicated degradations of VSR in real-world scenarios.
Compared with previous work, the new contributions lie in the following aspects:
1) We further explore the local and global frequency relations in frequency attention, and propose novel enhanced dual frequency attention (DFA), which is introduced in Section \ref{frequecny_attention}. The novel DFA can capture local and global frequency relations for frequency Transformers, which achieves better results on multiple VSR benchmarks.
2) We extend FTVSR to compressed videos with multiple compression algorithms in Section \ref{sec_eval_compress}, including Constant Rate Factor (CRF), Constant Quantization Parameters, and Constant Bitrate, which shows the stronger generalization ability of FTVSR.
3) We study the more complicated degradation in more real-world VSR, including Blur-based VSR in Section \ref{sec_blur}, Noise-based VSR in Section \ref{sec_noise}, and Real-world VSR in Section \ref{sec_real_world}.
4) The new FTVSR with DFA achieves state-of-the-art results on uncompressed, compressed, Blur-based, Noise-based, and real-world VSR benchmarks.

\section{Related Work}
% In this section, we will revisit the related works about video super-resolution and frequency learning. Most recent VSR methods enhance the uncompressed low-resolution videos, and is with know degradation kernels. Towards the real-world application, other methods study the compression problems and real-world video super-resolution. We will review the related works about VSR in these three aspects.

\subsection{Video Super-Resolution}
\subsubsection{Uncompressed Video Super-Resolution}
The VSR methods are developed from the image super-resolution method~\cite{dong2015image}. 
Existing video super-resolution approaches~\cite{chan2021basicvsr,yi2021omniscient,kim20183dsrnet,tian2020tdan,wang2019edvr,yang2020learning,sajjadi2018frame,liu2022learning} focus on restoring HR frames by extracting more temporal information, including the recurrent structure and sliding-window structure. 
For sliding-window structures, these methods~\cite{kim20183dsrnet,tian2020tdan,wang2019edvr} usually recover HR frames from adjacent LR frames within a sliding window. To align the temporal features, 3D convolution~\cite{kim20183dsrnet}, optical flow~\cite{kim2018spatio,tao2017detail} and deformable convolution~\cite{tian2020tdan,wang2019edvr} are widely-used.
Typically, 3DSRNet~\cite{kim20183dsrnet} adopts 3D convolution to extract temporal features. 
STTN~\cite{kim2018spatio} estimates optical flow in both spatial and temporal dimensions and uses the estimated flow to warp target frames. 
EDVR~\cite{wang2019edvr} adopts deformable convolution to align temporal features from adjacent frames. 
However, long-distance temporal features can not be utilized by these approaches. 

Other methods~\cite{tao2017detail,haris2019recurrent,isobe2020video,yi2021omniscient,chan2021basicvsr} based recurrent structure usually transmit long-distance temporal information from video frames by hidden state.
FRVSR~\cite{tao2017detail} restores the target SR frame with the hidden information from previously restored SR frames. 
RSDN~\cite{isobe2020video} designs a two-stream structure-detail block to learn textures by recurrent structure.
The bidirectional recurrent structure that fuses forward and backward propagation features in BasicVSR and IconVSR~\cite{chan2021basicvsr} brings notable benefits.
Although they aggregate the information from the whole sequence by hidden state, this mechanism loses the long-term modeling capabilities since the vanishing gradient.

Recently, with Transformer-based approaches~\cite{zeng2021improving,li2020mucan,cao2021video} are applied in VSR, great successes are achieved by using different attention~\cite{fu2017look,zheng2017learning} to capture temporal features. But they just can aggregate information from a few adjacent frames since the limited computational costs. Despite the impressive advancements made by these techniques, they primarily target uncompressed videos with known degradation processes and usually fail to recover the HR frames from compressed videos, or LR videos with blur degradation kernels and additive noises. In this work, we investigate the challenges of low-quality LR videos and propose a frequency-Transformer to handle these challenges.

\subsubsection{Compressed Video Super-Resolution}
Compressed video super-resolution is more challenging than uncompressed VSR because video compression results in losing information and carries out additional high-frequency artifacts.
Video denoising, training existing VSR models on compressed data, and designing a specific model for compression are three mainstream solutions to tackle the compression challenge.
For video denoising, COMISR~\cite{li2021comisr} applies existing video denoising~\cite{lu2018deep,lu2019deep,xu2019non} models on compressed LR videos, aiming to obtain clean LR videos by removing the compression artifacts. Then, they use SOTA VSR approaches~\cite{wang2019edvr,li2020mucan,chan2021basicvsr} to recover HR video from the denoised LR videos. 
Experimental results in \cite{li2021comisr} have demonstrated that the pre-cleaning method fails to recover HR details since the huge differences between the degradation kernel of VSR and the denoising model.
Moreover, COMISR~\cite{li2021comisr} has shown that joint training on compressed videos and uncompressed videos can not bring obvious gains, and may even hurt the the performance of VSR model without the specific model designs for compression. Thus, COMISR~\cite{li2021comisr} proposes a detail-aware module and a Laplacian module, which align high-resolution features and enhance HR frames, respectively. 
However, high-frequency textures can not be distinguished from artifacts by these designs since these signals are coupled in RGB images.
In this work, we will study the artifacts caused by different compression algorithms and tackle these challenges.

\subsubsection{Real-World Video Super-Resolution}
Towards real-world scenes, more complicated degradation on low-resolution videos brings huge challenges for VSR, such as different degradation kernels and additive noises.

To handle the real-world VSR challenges, a natural solution is that train the VSR model on real-world data. However, it's difficult to obtain the HR videos of the real-world LR videos. For example, recent RealVSR~\cite{yang2021real} captures paired data by a two-camera system. Thus, the degradation in the RealVSR dataset is specifically for their camera system. While other datasets~\cite{liu2013bayesian,nah2019ntire,xue2019video} are collected by downsampling on HR videos with pre-defined degradations, which limits the generalization ability of VSR methods.

To alleviate these problems, blind VSR~\cite{liu2013bayesian,lee2021dynavsr,pan2021deep} assumes that the LR videos are generated by applying a know degradation process but with unknown degradation parameters. The blind video super-resolution methods are developed from blind image super-resolution approaches~\cite{gu2019blind,hui2021learning,liang2021mutual,liang2021flow,xu2019towards,liu2022blind}.
These methods usually first estimate the degradation kernels and noise parameters, then design different modules to remove blur and noises. However, these particular designs for blur and noise have limitations on the variation of degradations and other real-world problems (e.g. compression).

Recent works~\cite{wang2021real,zhang2021designing} utilize generative adversarial network~\cite{wang2018esrgan} or data augmentation with more diverse degradations~\cite{wang2021real,zhang2021designing}. Although promising results have been achieved on real-world images, these image-level data augments are not feasible on videos~\cite{chan2022investigating}. While RealBasicVSR~\cite{chan2022investigating} firstly adopts pre-process module to clean the input frames, then applies the modern VSR methods~\cite{chan2021basicvsr,liu2022learning,chan2022basicvsr++} on the cleaned inputs. However, these pre-process methods still have generalizability limitations on complicated real-world degradations.
In this work, we propose a frequency Transformer and try to utilize the attention mechanism in the frequency domain to alleviate these challenges.

\subsection{Frequency Learning}
Existing studies of learning in the frequency domain can be divided into low-level restoration tasks~\cite{wang2016d3,ehrlich2020quantization,li2021learning,fritsche2019frequency} and high-level semantic tasks~\cite{ehrlich2019deep,xu2020learning,qin2021fcanet}. 
For high-level semantic tasks, existing methods usually transform images into the frequency domain to reduce the computational cost. 
FcaNet~\cite{qin2021fcanet} proposes frequency channel attention, in particular, to boost ResNet's performance on the classification tasks.
Numerous low-level studies investigate how to recover content details from a standpoint of frequency decomposition.
Some of them~\cite{fritsche2019frequency,li2021learning} design multi-branch networks to decompose features into different frequency bands. 
Typically, OR-Net~\cite{li2021learning} separates different frequency components by multi-branch CNNs and designs frequency enhancement units to enhance deep features. 
Other methods~\cite{wang2016d3,ehrlich2020quantization} transform images into the frequency domain. For example, D$^3$~\cite{wang2016d3} tries to remove JPEG compression 
 artifacts by a dual-domain restoration module. 
Moreover, Ehrlich \textit{et al.}~\cite{ehrlich2020quantization} propose a Y-channel correction network and color channel correction network to correct the JPEG artifacts. Thus, transferring RGB images into the frequency domain is beneficial to distinguish noises and texture, while existing VSR methods mainly focus on developing models in the pixel domain. In this work, inspired by the fact of that compression artifacts are generated in the process of quantification in the DCT domain, we introduce a frequency-Transformer to tackle the complicated degradation problems in video super-resolution.

\section{Method}

\subsection{Problem Formulation} 
The goal of VSR is to restore HR frames from their LR counterparts while existing frameworks do not take into account degradations such as blur, noise, as well as compression. 
Let $I_{LR} = \{I^t_{LR}|t\in[1,T]\}$ be a LR sequence of height $H$, width $W$, and frame length $T$. 
The corresponding HR frames are denoted as $I_{HR} = \{I^t_{HR}|t\in[1,T]\}$. Usually, the LR frames are generated by bicubic interpolation on HR frames. For complicated degradations, low-quality LR frame $I^t_{LR}$ can be generated as 
\begin{equation}
    I^t_{LR} = V((I^{t}_{HR}\otimes k)\downarrow_s +\mathcal{N}),
\end{equation}
where $\otimes$ means convolution and $k$ is the blur kernel. $\downarrow_s$ represents the downsampling with scale $s$. $\mathcal{N}$ means additive noises. For more complicated real-world scenes, we use $V(\cdot)$ to represent the video compression, and the quantization in the compression process brings artifacts on LR frames.
The restored super-resolution frames are denoted as $I_{SR} = \{I^t_{SR}|t\in[1,T]\}$ of height $\alpha H$, width $\alpha W$, in which
$\alpha$ represents the upsampling scale factor. 
% In this section, we first introduce the formulation of DCT-based visual tokens in Section \ref{dct_tokens}, and then introduce the proposed Frequency Transformer for compressed Video Super-Resolution (FTVSR) in Section \ref{frequecny_attention}. Finally, we introduce the recurrent structure for FTVSR in Section \ref{recurrent_FT_VSR}.

\subsection{Frequency-based Tokenization}
\label{dct_tokens}
To solve the problem of complicated degradations in video super-resolution, we propose a frequency-based patch representation. Following the previous works~\cite{gueguen2018faster,ehrlich2019deep,ehrlich2020quantization}, we adopt the widely-used method, Discrete Cosine Transform (DCT), to transfer an image into the frequency domain.
 
\begin{figure*}
\centering
\includegraphics[width=\textwidth]{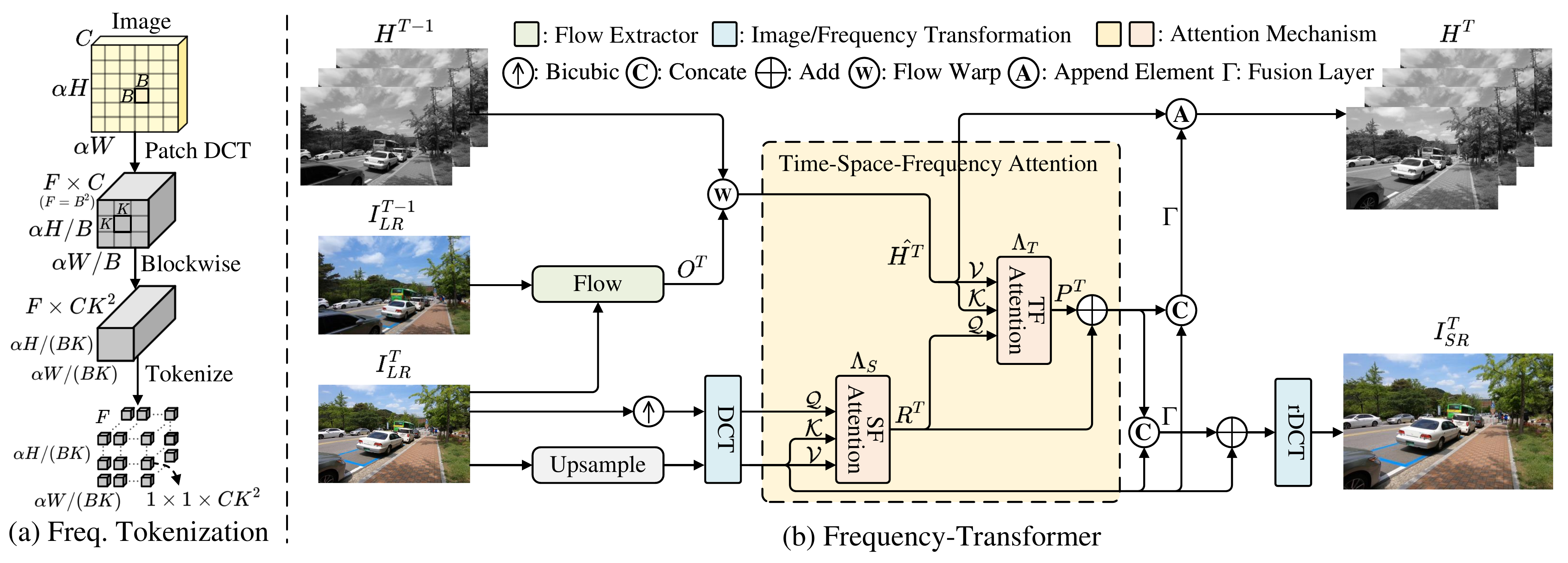}
% \vspace{-0.6cm}
\caption{(a) The process of DCT-based frequency tokenization. (b) The illustration of a Frequency-Transformer with attention $\Lambda_{ST}$ in a joint time-space-frequency domain. 
Given an LR video, Frequency-Transformer extracts frequency tokens and computes time-space-frequency (TSF) attention on these tokens.
TSF includes Space-Frequency attention $\Lambda_S$ and Time-Frequency attention $\Lambda_T$. 
% The $Q$, $K$, $V$ of $\Lambda_S$ are tokens from videos frame sampled $I^T_{LR}$ by bicubic and upsample network, respectively. $R^T$ is the output of $\Lambda_S$, further sums with hidden states $\Hat{H}^T$ warped from past hidden states by flow $O^T$, as the $K$ and $V$ of $\Lambda_T$ attention. $P^T$ is the output of $\Lambda_T$, which is further used to update the hidden state and recover SR frame $I^T_{SR}$.
}
\label{fig:FTVSR}
\end{figure*}

\subsubsection{Discrete Cosine Transform} Discrete Cosine Transform (DCT) projects an image into a set of cosine components for different 2D frequencies. Given an image patch $P$ of height $B$ and width $B$, a $B \times B$ DCT block $D$ is generated as:
\begin{equation}
\label{eq_dct}
% \begin{aligned}
% D(u, v) =  c(u)c(v)\sum^{B-1}_{x=0}\sum^{B-1}_{y=0}P(x,y)cos[\frac{(2x+1)u\pi}{2B}]cos[\frac{(2y+1)v\pi}{2B}],
\begin{aligned}
D(u, v) & =  c(u)c(v)\sum^{B-1}_{x=0}\sum^{B-1}_{y=0}P(x,y)cos(a)cos(b),\\
s.t. ~a &=\frac{(2x+1)u\pi}{2B},~ b=\frac{(2y+1)v\pi}{2B},
\end{aligned}
\end{equation}
where $x$ and $y$ are the 2D indexes of pixels. $u\in [0,B-1]$ and $v\in [0,B-1]$ are the 2D indexes of frequencies. $c(\cdot)$ represents normalizing scale factor to enforce orthonormality and $c(u) = \sqrt{\frac{1}{B}}$ if $u=0$, else $c(u) = \sqrt{\frac{2}{B}}$. The DCT and its inversion are denoted as $\text{DCT}(\cdot)$ and $\text{rDCT}(\cdot)$, respectively.

\subsubsection{DCT-based Frequency Tokenization}
Given an LR sequence, we firstly upsample the $I_{LR}$ by an upsampling network $\varphi(\cdot)$. 
For each frame, we transform each channel of RGB image into the frequency domain by applying DCT on the patches of shape $B\times B$ as Equation \ref{eq_dct}, which can be formulated as: 
\begin{equation}
    D_{LR}(u,v) = \text{DCT}(\varphi (I_{LR})),
\end{equation}
where $D_{LR}(u,v)$ of shape $T\times F\times C\times \frac{\alpha H}{B}\times \frac{\alpha W}{B}$ represents the transformed 2D spectral map from LR image. $T$, $F$, $C$, $\frac{\alpha H}{B}$ and $\frac{\alpha W}{B}$ represent sequence length, frequency dimensions, image channels, height, and width, respectively. The frequency number is $F = B^2$. 

For a spectral frame $D_{LR}(u,v)$, we split the frequency dimension to form $F$ visual tokens. The frequency tokens set $\mathcal{T}$ can be represented as: 
\begin{equation}
    \mathcal{T} = \{\tau_f, f\in [1, F]\},
\end{equation}
where $\tau_f$ represents the frequency token in $f^{th}$ frequency, which has a feature size of $C\times \frac{\alpha H}{B}\times \frac{\alpha W}{B}$. 
This frequency tokenization mechanism carries out the information exchange between different frequency bands and forces the neural network to ``fairly" treat low-frequency signals and high-frequency signals, which is beneficial to preserve high-frequency visual details. Combined with the frequency attention mechanism in Section \ref{frequecny_attention}, low-frequency information (e.g. object structure) can guide the network to restore the high-frequency textures.

To capture the interactions of frequency signals among different spatial blocks, the spectral maps are split into a set of blocks. The block size is $K\times K$. 
To further utilize temporal information in the video, the same tokenization process is applied to all video frames to extract temporal frequency tokens.
Thus, the fine-grained frequency tokens in both spatial and temporal dimensions are generated, which can be denoted as:
\begin{equation}
    \mathcal{T} = \{\tau_{(t,i,f)}, t\in [1,T], i\in [1,N], f\in [1, F]\},
\end{equation}
where each token $\tau_{(t,i,f)}$ has a shape of $C\times K\times K$. $N$ represents the generated block number in each frame. Different from traditional vision Transformers~\cite{dosovitskiy2020image,liu2021swin,cao2021video}, which crop image patches and form a set of spatial visual tokens, our tokens are based on different frequency bands. In a nutshell, we generate $N$ blocks for each spectral frame $D^t_{LR}$, and each block has  DCT-based frequency tokens $\tau$ of the number of $F$. The total number of frequency tokens is $T\times N \times F$. Figure \ref{fig:FTVSR} (a) presents more details about the whole tokenization process.

\subsection{Frequency-based Attention}
\label{frequecny_attention}
The inputs of the frequency Transformer are DCT-based visual tokens, which have been generated in Section \ref{dct_tokens}. To better take advantage of temporal information for VSR, the query tokens $\mathcal{Q}$ are extracted from spectral map $D^T_{LR}$. Keys $\mathcal{K}$ and values $\mathcal{V}$ are extracted from spectral maps $\{D^t_{LR}, t\in [1, T-1]\}$. For a target frame $D^T_{LR}$, the query, key, and value sets are denoted as:
\begin{equation}
\begin{aligned}
    \mathcal{Q} & = \{ \tau^q_{(T,i,f)}, i\in [1,N], f\in [1,F] \}, \\
    \mathcal{K} & = \{ \tau_{(t,i,f)}^k, t\in [1,T-1],i\in [1,N], f\in [1,F]\}, \\
    \mathcal{V} & = \{ \tau_{(t,i,f)}^v, t\in [1,T-1],i\in [1,N], f\in [1,F]\},
\end{aligned}
\end{equation}
where $\tau_{(T,i,f)}^q$, $\tau_{(t,i,f)}^k$, and $\tau_{(t,i,f)}^v$ represent the query, key, and value tokens, respectively. 
Each token is extracted from spectral maps among time, space, and frequency dimensions according to the needs of computing different kinds of frequency attention.

In the following, we first introduce basic frequency attention (FA).
To handle the different types of degradation (e.g. blur, noise, and compression), we propose enhanced dual frequency attention (DFA). 
FA and DFA are the two essential components of self-attention in the frequency domain.
To explore different self-attention schemes for video processing in the frequency domain, we study how to use frequency and dual frequency attention in spatial and temporal dimensions. Based on FA and DFA, we will introduce Space-Frequency attention (SF), Time-Frequency attention (TF), and joint Time-Space-Frequency (TSF) attention.

\subsubsection{Frequency Attention}
The frequency attention aims to capture the relationship between different frequency bands. The structure of frequency attention is shown in Figure \ref{fig:fa_dfa} (a). Since each frequency band contains the global information of the whole image, the basic frequency attention is Global Frequency Attention (GFA).
Given a query token $\tau^q_{f}$ at the $f^{th}$ frequency, the uniform formulation of frequency attention (denoted as $\Lambda$) is:
\begin{equation}
\label{eq_fa}
    \Lambda (\tau^q_{f}, \tau_{\hat{f}}^k, \tau_{\hat{f}}^v) = \text{SM}(\frac{\tau_{f}^q \cdot \tau_{\hat{f}}^k}{\sqrt{d^k}})\tau_{\hat{f}}^v, \hat{f} \in [1,F],
\end{equation}
where $\text{SM}$ represents the softmax activation function and $d^k$ denotes the normalization factor. Note that there is a feed-forward network (FFN) after frequency attention, which is omitted in this paper.

Based on equation \ref{eq_fa}, the multi-head frequency attention can be represented as
\begin{equation}
\label{eq_mfa}
\begin{aligned}
\Lambda^M(\mathcal{Q}, \mathcal{K}, \mathcal{V}) &= \text{Concat}(head_1, ..., head_H), \\
s.t. ~head_h &= \Lambda (\mathcal{Q}_h, \mathcal{K}_h, \mathcal{V}_h), h\in [1,H],
\end{aligned}
\end{equation}
where $H$ means head number. $\text{Concat}(\cdot)$ means the operation of concatenating matrix.

After transferring images into the frequency domain, the DCT-based representation is beneficial to recover high-frequency visual details. Moreover, frequency attention can capture the interactions among different frequency bands, which guides the generation of high-frequency textures with the guidance of low-frequency information. 

\subsubsection{Dual Frequency Attention}
For low-quality video super-resolution, the high degradation (e.g. blur, noise, and compression) brings huge challenges for current VSR methods. For example, blur degradation and additive noises have similar patterns on whole images, which need global information to handle. However, compression artifacts show different patterns on each patch of the images since the different quantization errors on the patches in the process of compression, need to focus on local information to handle.

\begin{figure}[!t]
\centering
\includegraphics[width=\columnwidth]{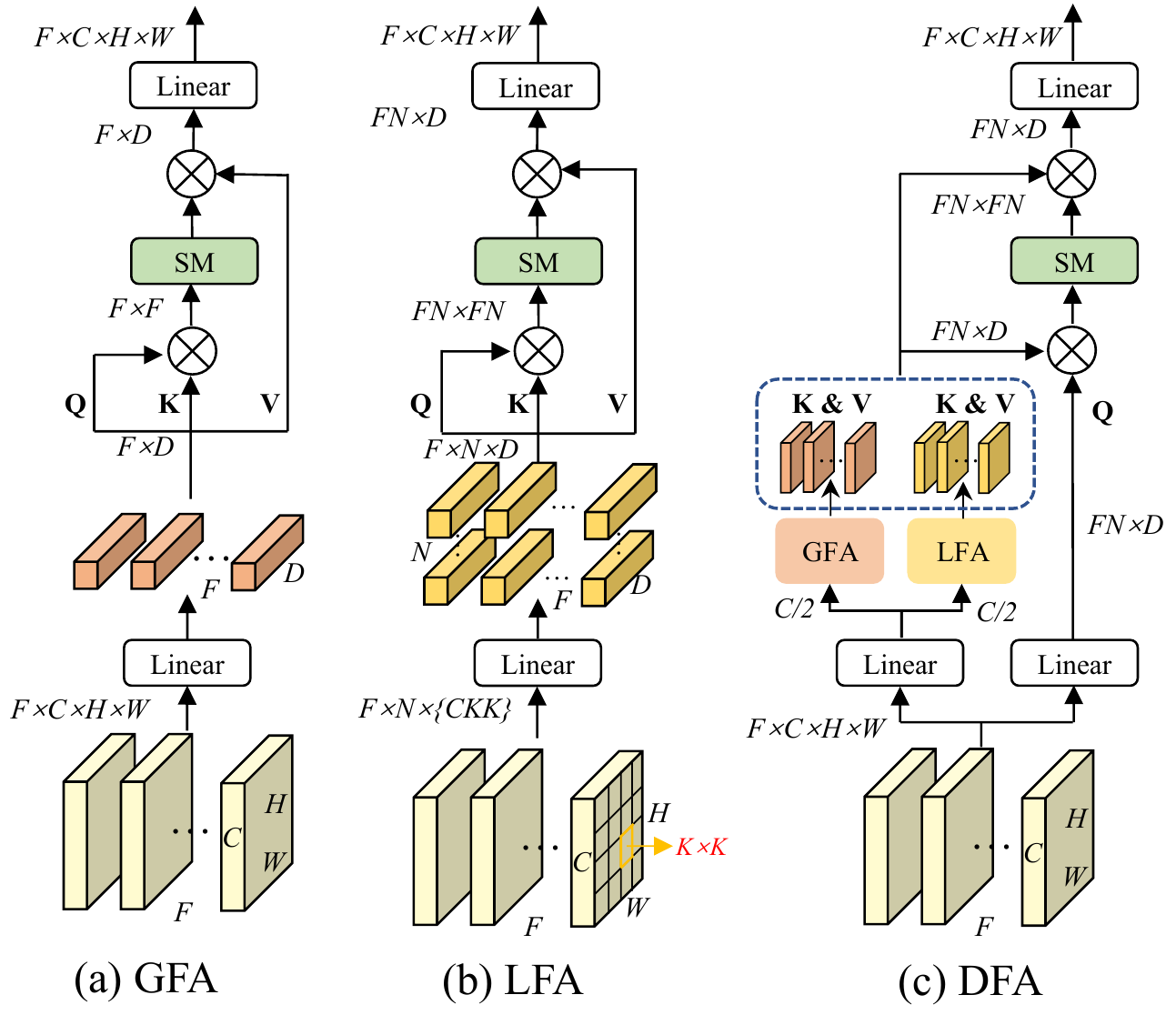}
% \vspace{-0.6cm}
\caption{The illustration of (a) Global Frequency Attention, (b) Local Frequency Attention, (c) Dual Frequency Attention. $\otimes$ means matrix multiplication and $SM$ means $softmax$ function. }
\label{fig:fa_dfa}
\end{figure}

To capture the local information in the frequency domain, we propose Local Frequency Attention (LFA), represented as $\Lambda_l$, which computes the frequency relationship among different spatial patches. The structure of LFA is shown in Figure \ref{fig:fa_dfa} (b). LFA adopts a similar architecture as Equation \ref{eq_fa} in global frequency attention. Different from GFA, LFA extracts frequency tokens from each patch of size $K\times K$, thus generating $F\times N$ frequency tokens, where $N$ means the patch number. Following equation \ref{eq_fa} and \ref{eq_mfa}, LFA computes a frequency attention matrix of size $FN\times FN$ and restores tokens by performing matrix multiplication of attention matrix and value tokens. 

To tackle the multiple complicated degradations in VSR, we propose Dual Frequency Attention (DFA), which can capture the global and local information in the frequency domain. Dual frequency attention can handle more complicated degradation in the real-world video super-resolution by combining GFA and LFA. The structure of dual frequency attention is shown in Figure \ref{fig:fa_dfa} (c). The dual frequency attention $\widetilde{\Lambda}(\mathcal{Q}, \mathcal{K}, \mathcal{V})$ can be represented as
\begin{equation}
\label{eq_dfa}
    \widetilde{\Lambda} =\text{SM}(\frac{\mathcal{Q}\cdot (\Lambda_g(\mathcal{K}) \oplus \Lambda_l(\mathcal{K}))}{\sqrt{d^k}}) (\Lambda_g(\mathcal{V}) \oplus \Lambda_l(\mathcal{V}))
\end{equation}
where $\Lambda_g$ and $\Lambda_l$ represent global frequency attention and local frequency attention, respectively. $\oplus$ means concat operation. Following equation \ref{eq_mfa} and based on equation \ref{eq_dfa}, the multi-head dual frequency attention can be represented as $\widetilde{\Lambda}^M(\mathcal{Q},\mathcal{K}, \mathcal{V})$.

Dual frequency attention firstly splits tokens into two groups along feature dimensions, which are sent into GFA and LFA to learn global and local frequency relations, respectively. Then, the outputted tokens by GFA and LFA are concatenated as key and value, which further are used to compute frequency attention with original query tokens. 

In dual frequency attention, the GFA branch captures global information, so that global degradation (e.g. blur and additive noise) can be mitigated and generate high-frequency texture. LFA branch can capture the local frequency relations inside a patch and among other spatial patches, which is beneficial to handle local degradation (e.g. compression artifacts). DFA aggregates local and global information in the frequency 
domain, which can greatly improve the model's ability to deal with complex and different degradations in the real-world VSR.

\subsection{Frequency Attention in Video}
In Section \ref{frequecny_attention}, we introduce the essential frequency attentions: global frequency attention $\Lambda_g$, local frequency attention $\Lambda_l$, and dual frequency attention $\widetilde{\Lambda}$. To expand frequency attention into video processing, we will study frequency attention in a joint time-space-frequency domain. In the following, we first introduce frequency attention by taking spatial or temporal dimension into account, then we will explore the frequency attention in the joint time-space-frequency dimension.

\begin{figure*}[!t]
\centering
\includegraphics[width=0.9\textwidth]{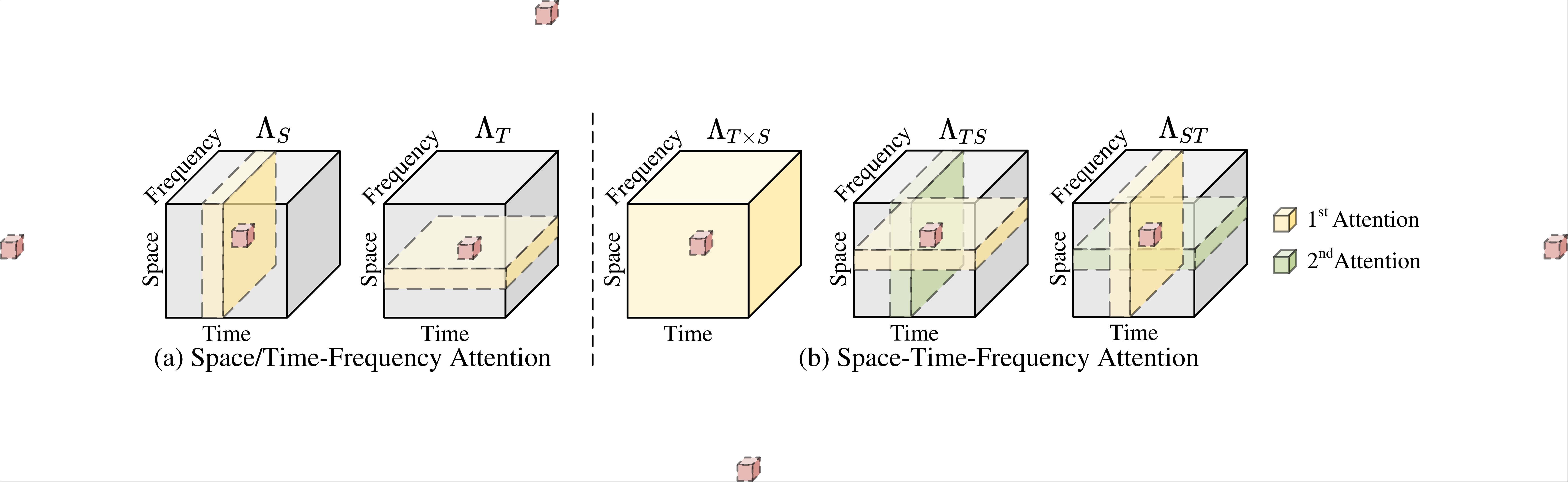}
% \vspace{-0.6cm}
\caption{(a) The illustration of space/time-frequency attention. (b) The illustration of space-time-frequency attention. 
The red cube indicates the query token. The yellow and green areas mean the candidate areas for computing attention with query following the order of yellow first and then green.
}
\label{fig:freqAttention}
\end{figure*}

\subsubsection{Space/Time-Frequency Attention}
Space-Frequency (SF) attention computes the frequency attention weights between spatial blocks. The visualization of SF is shown in Figure \ref{fig:freqAttention} (a). 
For a query token $\tau^q_{(i,f)}$ at the $f^{th}$ frequency in the $i^{th}$ block, the SF attention is $\Lambda_S(\tau^q_{(i,f)}, \tau^k_{(\hat{i},\hat{f})}, \tau^v_{(\hat{i},\hat{f})}), \hat{i}\in [1,N], \hat{f}\in [1,F]$, which computes the frequency attention as Equation \ref{eq_fa} in spatial dimension. The inputs of $\Lambda_S$ are space-frequency tokens $\tau_{(i,f)}$. Since the tokens are extracted from both spatial and temporal dimensions, $N\times F$ tokens are generated as inputs to compute SF attention. The SF attention computes the local frequency relations in a single space dimension, thus it can be regarded as that local frequency attention $\Lambda_l$ is performed inside an image.

The Time-Frequency (TF) attention is computed on the blocks with the same spatial position from different video frames. The visualization of TF attention is shown in Figure \ref{fig:freqAttention} (a). 
Given a query token $\tau^q_{(t,f)}$, the TF attention is $\Lambda_T(\tau^q_{(t,f)}, \tau^k_{(\hat{t}, \hat{f})}, \tau^v_{(\hat{t}, \hat{f})}), \hat{t}\in [1,T-1],\hat{f}\in [1,F]$, which computes frequency attention as Equation \ref{eq_fa} in temporal dimension. The inputs of $\Lambda_T$ are time-frequency tokens $\tau_{(t,f)}$. Since the tokens are extracted from both time and frequency dimensions, $T\times F$ tokens are generated as inputs to compute TF attention. The TF attention computes the local frequency relations among the same spatial patches in different temporal frames, thus it can be regarded as that local frequency attention $\Lambda_l$ is performed on the local patches among different temporal frames.

\subsubsection{Time-Space-Frequency Attention}

For VSR, spatio-temporal information is important to recover HR frames. To utilize spatio-temporal information in the frequency domain, we design Time-Space-Frequency (TSF) attention to explore how to apply frequency attention in a joint space-time-frequency domain. TSF attention is the combination of SF attention and TF attention. TSF attention can be categorized as joint SF and TF attention and divided SF and TF attention. The architectures of different types of TSF attention are shown in Figure \ref{fig:freqAttention} (b). 

Given a query token $\tau^q_{(t,i,f)}$, joint TSF attention $\Lambda_{T\times S}(\tau^q_{(t,i,f)}, \tau^k_{(\hat{t},\hat{i},\hat{f})}, \tau^v_{(\hat{t},\hat{i},\hat{f})}), $ $\hat{t}\in [1,T-1], \hat{i} \in [1,N],\hat{f}\in [1,F]$ computes the frequency attention in a joint space-time-frequency domain following Equation \ref{eq_fa}. 
The inputs of $\Lambda_{T\times S}$ are fine-grained tokens $\tau_{(t,i,f)}$ in a joint time-space-frequency domain. Thus, there are $T\times N \times F$ tokens extracted among time, space, and frequency dimensions for joint TSF attention.

For divided TSF attention, according to the order of computing $TF$ and $SF$ attention, two types of TSF ($\Lambda_{ST}$ and $\Lambda_{TS}$) are proposed. $\Lambda_{ST}$ can be represented as: 
\begin{equation}
\label{eq_ST}
\begin{aligned}
      \Lambda_{ST}(\tau^q_{(t,i,f)}, \tau^k_{(\hat{t},\hat{i},\hat{f})}, \tau^v_{(\hat{t},\hat{i},\hat{f})}) &= \Lambda_{T}(\hat{\tau}^q_{(t,f)}, \tau_{(\hat{t},\hat{f})}, \tau_{(\hat{t},\hat{f})}), \\
      s.t. ~\hat{\tau} &= \Lambda_{S}(\tau^q_{(i,f)}, \tau^k_{(\hat{i},\hat{f})},\tau^k_{(\hat{i},\hat{f})}),
\end{aligned}
\end{equation}
where $\hat{t}  \in [1,T-1], \hat{i}\in [1,N], \hat{f}\in [1,F]$. 
The divided TSF attention $\Lambda_{ST}$ represents that computing space-frequency attention $\Lambda_S$ firstly, then computing time-frequency attention $\Lambda_T$. The architecture of $\Lambda_{ST}$ is shown in Figure \ref{fig:freqAttention} (b).

Another divided TSF attention $\Lambda_{TS}$ can be formulated as $\Lambda_{TS}(\tau^q_{(t,i,f)}, \tau^k_{(\hat{t},\hat{i},\hat{f})}, \tau^v_{(\hat{t},\hat{i},\hat{f})})$.
The divided TSF attention $\Lambda_{TS}$ represents the attention that computes time-frequency attention $\Lambda_{T}$ firstly, then computes space-frequency attention $\Lambda_{S}$. The computing process of $\Lambda_{TS}$ is similar to Equation \ref{eq_ST} and its structure is shown in Figure \ref{fig:freqAttention} (b).

In our experiments, $\Lambda_{ST}$ performs best for the frequency Transformer. This is because, in VSR, degraded frames should be first restored by the space-frequency attention then the recovered textures could be used to benefit temporal learning in the time-frequency attention.

To handle the complicated degradations in the real-world VSR, the basic frequency attention module ($\Lambda_g$ or $\Lambda_l$) in TSF attention can be replaced by enhanced dual frequency attention $\widetilde{\Lambda}$.

\subsection{Frequency Transformer}
To recover HR videos, we use a similar recurrent structure as TTVSR~\cite{liu2022learning}. Each HR frame is restored from its LR counterparts and a propagation hidden state $H$. Given a LR frame $I^T_{LR}$, the SR frame can be restored as:
\begin{equation}
% \label{eq_Isr}
\begin{aligned}
    I^T_{SR} & = \text{rDCT}(T_{freq}(\mathcal{Q}, \mathcal{K}, \mathcal{V})) \\
    & = \text{rDCT}(\Gamma(A_{freq}(\mathcal{Q}, \mathcal{K}, \mathcal{V}), D_{LR}^T)+ D_{LR}^T),
\end{aligned}
\end{equation}
where $T_{freq}$ represents the Frequency Transformer. $A_{freq}$ represents the frequency attention used in $T_{freq}$ and $A_{freq} \in \{\Lambda_S, \Lambda_T, \Lambda_{T\times S}, \Lambda_{TS}, \Lambda_{ST}\}$. $\Gamma$ represents the fusion operation which concatenates the outputs of $A_{freq}$ and $D^T_{LR}$, then reduces the dimensions of the concatenated features by Linear layer.

For example, a frequency Transformer formed by divided TSF attention $\Lambda_{SF}$ is shown in Figure \ref{fig:FTVSR} (b). The output $P^T$ of TSF can be formulated as:
\begin{equation}
\begin{aligned}
P^T &~= \Lambda_T(R_T, \Hat{H}^T, \Hat{H}^T),\\
s.t. ~R^T &~= \Lambda_S(\mathcal{Q}_S, \mathcal{K}_S,  \mathcal{V}_S), \Hat{H}^T = W(H^{T-1}, O^T).
\end{aligned}
\end{equation}
$\Hat{H}^T$ represents the hidden states warped from past frames $H^{T-1}$ according to flow $O^T$. $W$ represents the flow warp operation as \cite{chan2021basicvsr}. $H^T$ is updated by the output $P^T$ of TSF attention and the DCT-based features $D^T_{LR}$. $\mathcal{Q}_S$ are extracted from upsampled $I^T_{LR}$ by Bicubic upsampling while $\mathcal{K}$ and $\mathcal{V}$ are extracted from upsampled $I^T_{LR}$ by a upsample neural network. The difference between upsample operations brings the location guidance of the hard-to-recover parts, which should pay more attention to it. $\mathcal{Q}_T$ is the temporal-frequency query tokens for $\Lambda_T$. The output $P^T$ of TSF attention $\Lambda_{SF}$ is used to recover SR frames, which can be formulated as:
\begin{equation}
\label{eq_final_ftvsr}
  I^T_{SR}  = \text{rDCT}(\Gamma(P^T + R^T,D_{LR}^T) + D_{LR}^T).
\end{equation}

\begin{table*}[!t]
  \caption{Evaluation on the \textbf{uncompressed} videos of REDS4~\cite{nah2019ntire} and Vid4~\cite{xue2019video} for $4\times$ VSR. Each entry shows the PSNR$\uparrow$/SSIM$\uparrow$.}
%   \vspace{-0.2cm}
\tiny
  \centering
  \renewcommand\arraystretch{1.2}
  \renewcommand\tabcolsep{4pt}
  \resizebox{\textwidth}{!}{
  \begin{tabular}{ l|c|c|c|c|c|c|c|c}
    \hline
    
    \hline
    Methods & TOFlow\cite{xue2019video} & DUF\cite{jo2018deep}  & EDVR\cite{wang2019edvr} & COMISR\cite{li2021comisr} & BasicVSR\cite{chan2021basicvsr} & IconVSR\cite{chan2021basicvsr} & TTVSR\cite{liu2022learning} & FTVSR\\
    \hline
    
    \hline
    REDS4~\cite{nah2019ntire} & 27.98/0.799 & 28.63/0.825 & 31.09/0.880 & 29.68/0.868 & 31.42/0.890 & 31.67/0.895 & 32.12/0.901 & \textbf{32.42/0.907} \\
    \hline
    Vid4~\cite{xue2019video} & 25.85/0.766 & 27.38/0.832 & 27.85/0.850 & 27.31/0.840 & 27.96/0.855 & 28.04/0.857 & 28.40/0.864 & \textbf{28.70/0.869}\\
    \hline
    
    \hline
  \end{tabular}
  }
  \label{tab_reds_unc}
%   \vspace{-0.5cm}
\end{table*}

% During training, the inputs are LR videos $I_{LR}$ and outputs are SR videos $I_{SR}$. We use
Following previous works~\cite{chan2021basicvsr,li2021comisr}, we apply Charbonnier penalty loss~\cite{lai2017deep} on each video frames. The total loss $\mathcal{L}$ is the average of frames, 
\begin{equation}
    \mathcal{L} = \frac{1}{T}\sum_{t=1}^T\sqrt{||I^t_{HR} - I^t_{SR}||^2 + \epsilon^2},
\end{equation}
where $\epsilon$ is a constant value and $\epsilon =1e-3$.

\section{Experiments}
% In this section, we first our implementation details. We then introduce the benchmarks and the new evaluation metrics for compressed VSR. Next, we show our state-of-the-art results on two benchmarks and on the real-world compressed videos. Finally, we show the ablation study of FTVSR.

\begin{table*}[t]
  \caption{Evaluation and comparison with state-of-the-art methods on the \textbf{compressed} videos from REDS4~\cite{nah2019ntire} dataset. Following previous works, each entry shows the PSNR$\uparrow$/SSIM$\uparrow$ on RGB channels. 
  % \textcolor{red}{Red} indicates the best results and \textcolor{blue}{{blue}} indicates the second best performances.
  }
%   \vspace{-0.2cm}
\tiny
  \centering
  \renewcommand\arraystretch{1.2}
  \renewcommand\tabcolsep{6pt}
  \resizebox{\textwidth}{!}{
  \begin{tabular}{ l| c | c | c | c | c | c | c}
    \hline
    
    \hline
    \multirow{2}{*}{Methods} &\multicolumn{3}{c|}{Average of clips with Compression} &\multicolumn{4}{c}{Per clip with Compression CRF25} \\
    \cline{2-8}
      & CRF15 & CRF25 & CRF35 &  Clip\_000 &  Clip\_011 &  Clip\_015 &  Clip\_020  \\
    \hline
    DUF~\cite{jo2018deep} & 25.61/0.775 & 24.19/0.692 & 22.17/0.588 & 23.46/0.622 & 24.02/0.686 & 25.76/0.773 & 23.54/0.689  \\
    EDVR~\cite{wang2019edvr} & 28.72/0.805 & 25.98/0.706 & 23.36/0.600 & 24.38/0.629 & 26.01/0.702 & 28.30/0.783 & 25.21/0.708 \\
    TecoGan~\cite{chu2020learning} & 26.93/0.768 & 25.46/0.690 & 22.95/0.589 & 24.01/0.624 & 25.39/0.682 & 27.95/0.768 & 24.48/0.686 \\
    FRVSR~\cite{sajjadi2018frame} & 27.61/0.784 & 25.72/0.696 & 23.22/0.579 & 24.25/0.631 & 25.65/0.687 & 28.17/0.770 & 24.79/0.694  \\
    RSDN~\cite{isobe2020video} & 27.66/0.768 & 25.48/0.679 & 23.03/0.579 & 24.04/0.602 & 25.40/0.673 & 27.93/0.766 & 24.54/0.676 \\

    % VSR-T~\cite{cao2021video} & r & & & & & & & \\
    BasicVSR~\cite{chan2021basicvsr} & 29.05/0.814 & 25.93/0.704 & 23.22/0.596 & 24.37/0.628 & 26.01/0.702 & 28.13/0.777&25.21/0.709 \\
    IconVSR ~\cite{chan2021basicvsr} &29.10/0.816 &25.93/0.704 & 23.22/0.596 &24.35/0.627 & 26.00/0.702 & 28.16/0.777& 25.22/0.709 \\
    MuCAN~\cite{li2020mucan} & 28.67/0.804 & 25.96/0.705 & 23.55/0.600 & 24.39/0.628 & 26.02/0.702 & 28.25/0.781 & 25.17/0.707  \\
    % \hline
    COMISR~\cite{li2021comisr} &
    28.40/0.809 & 
    26.47/0.728 & 23.56/0.599 & 24.76/0.660 & 26.54/0.722 & 29.14/0.805 & 25.44/0.724 \\
    \hline
    \textbf{FTVSR} & 
    \textBF{30.55/0.854} & 
    \textBF{28.13/0.778} & 
    \textBF{24.87/0.660} & 
    \textBF{26.12/0.703} & 
    \textBF{28.81/0.779} & 
    \textBF{30.21/0.839} & 
    \textBF{27.38/0.782} \\
    \hline
    
    \hline
  \end{tabular}
  }
  \label{tab_compress_reds}
\end{table*}

\subsection{Implementation Details}
We employ the Adam optimizer and the Cosine Annealing technique with $beta 1=0.9$ and $beta 2=0.99$ for training.
The batch size is 8 and the starting learning rate is $2\times 10^4$.
The training video length is 10 for the quick ablation study and 40 for the better final performance. 
The input size of the image for training FTVSR is $64\times 64$ and the super-resolution scale is $4\times$. 
Random rotations, vertical flips, and horizontal flips are used for data augmentations.
FTVSR is trained with 100k iterations for rapid ablation study and 400k iterations for the final model.
For quick evaluation, the backbone network for all ablation studies is BasicVSR~\cite{chan2021basicvsr} while the backbone of the final model is TTVSR~\cite{liu2022learning} for better results.
Following BasicVSR~\cite{chan2021basicvsr} and TTVSR~\cite{liu2022learning}, the weights of the motion estimation module are fixed in the first 5k iterations for stable training and then released. 
In order to maintain that the input images can be converted into spectral maps by using DCT, input images are padded before applying DCT.
% During inference, input images are padded with the edge values to keep that the images can be transformed into spectral maps by DCT. 
After model forwarding, the padding pixels are removed after applying rDCT to transform spectral maps into RGB images.
During inference, limited to the CUDA memory, the images are cropped into 4 patches.

\subsection{Datasets and Evaluation Metrics}
\subsubsection{Datasets}
To make fair comparison with exsiting methods~\cite{cao2021video,chan2021basicvsr}, REDS~\cite{nah2019ntire} and Vimeo-90K~\cite{xue2019video} datasets are used for training. The REDS dataset includes 270 videos and 100 frames with a resolution of $1280\times 720$ in each video. Following previous works~\cite{chan2021basicvsr,cao2021video,wang2019edvr,li2020mucan}, four sequences are used for evaluation, termed REDS4. There are 64,612 training videos in the Vimeo-90K dataset, in which each video contains 7 frames of image size $448\times 256$. For evaluation, 4 videos are used as previous works~\cite{chan2021basicvsr,li2021comisr}, termed Vid4. Each video in the Vid4 dataset contains $30$ to $50$ frames.

For real-world VSR evaluation, we use VideoLQ~\cite{chan2022investigating} dataset. The videos in VideoLQ are collected from real websites, such as Flickr and YouTube. The videos in VideoLQ contain various contents and image sizes, which cover more real-world degradations. Each video includes 100 frames without scene changes. 

\begin{table*}[t]
  \caption{Evaluation and comparison with state-of-the-art methods on the \textbf{compressed} videos from Vid4~\cite{xue2019video} dataset. Following previous works, each entry shows the PSNR$\uparrow$/SSIM$\uparrow$ on Y-channel. 
  % \textcolor{red}{Red} indicates the best results and \textcolor{blue}{{blue}} shows the second-best performances.
  }
%   \vspace{-0.2cm}
\tiny
  \centering
    \renewcommand\arraystretch{1.2}
  \renewcommand\tabcolsep{6pt}
  \resizebox{\textwidth}{!}{
  \begin{tabular}{ l| c | c | c | c | c | c | c}
    \hline
    
    \hline
    \multirow{2}{*}{Methods}& \multicolumn{3}{c|}{Average of clips with Compression} & \multicolumn{4}{c}{Per clip with Compression CRF25}\\
    \cline{2-8}
      & CRF15 & CRF25 & CRF35 & calendar & city & foliage & walk  \\
    \hline
    DUF~\cite{jo2018deep} & 24.40/0.773 & 23.06/0.660 & 21.27/0.515 & 21.16/0.634 & 23.78/0.632 & 22.97/0.603 & 24.33/0.771  \\
    EDVR~\cite{wang2019edvr} & 26.34/0.771 & 24.45/0.667 & 22.31/0.534 & 21.69/0.648 & 25.51/0.626 & 24.01/0.606 & 26.72/0.786 \\
    TecoGan~\cite{chu2020learning} & 25.25/0.741 & 23.94/0.639 & 21.99/0.479 & 21.34/0.624 & 25.26/0.561 & 23.50/0.592 & 25.73/0.756  \\
    FRVSR~\cite{sajjadi2018frame} & 26.01/0.766 & 24.33/0.655 & 22.05/0.482 & 21.55/0.631 & 25.40/0.575 & 24.11/0.625 & 26.21/0.764  \\
    RSDN~\cite{isobe2020video} & 26.58/0.781 & 24.06/0.650 & 21.29/0.483 & 21.72/0.650 & 25.28/0.615 & 23.69/0.591 & 25.57/0.747  \\
    BasicVSR~\cite{chan2021basicvsr} & 26.56/0.780 &24.28/0.656 & 21.97/0.509 & 21.64/0.641 & 25.45/0.620 &23.79/0.586 &26.26/0.774  \\
    IconVSR ~\cite{chan2021basicvsr} & 26.65/0.782 &24.31/0.657 &21.97/0.509 & 21.67/0.644 &25.46/0.621 &23.83/0.588 & 26.26/0.774 \\
    MuCAN~\cite{li2020mucan} & 25.85/0.753 & 24.34/0.661 & 22.26/0.531 & 21.60/0.643 & 25.38/0.620 & 23.93/0.599 & 26.43/0.782  \\
    % \hline
    COMISR~\cite{li2021comisr} & 
    26.43/0.791 & 
    24.97/0.701 & 
    22.35/0.509 & 
    22.81/0.695 & 
    25.94/0.640 & 
    24.66/0.656 & 
    26.95/0.799  \\
    \hline
    \textbf{FTVSR} & 
    \textBF{27.40/0.811} & 
    \textBF{25.38/0.706} & 
    \textBF{22.61/0.540} & 
    \textBF{22.97/0.720} &
    \textBF{26.29/0.670} & 
    \textBF{24.94/0.664} & 
    \textBF{27.30/0.816} \\
    \hline
    
    \hline
  \end{tabular}
  }
  \label{tab_com_vid4}
%   \vspace{-0.5cm}
\end{table*}

\begin{figure*}[]
\centering
\includegraphics[width=0.95\textwidth]{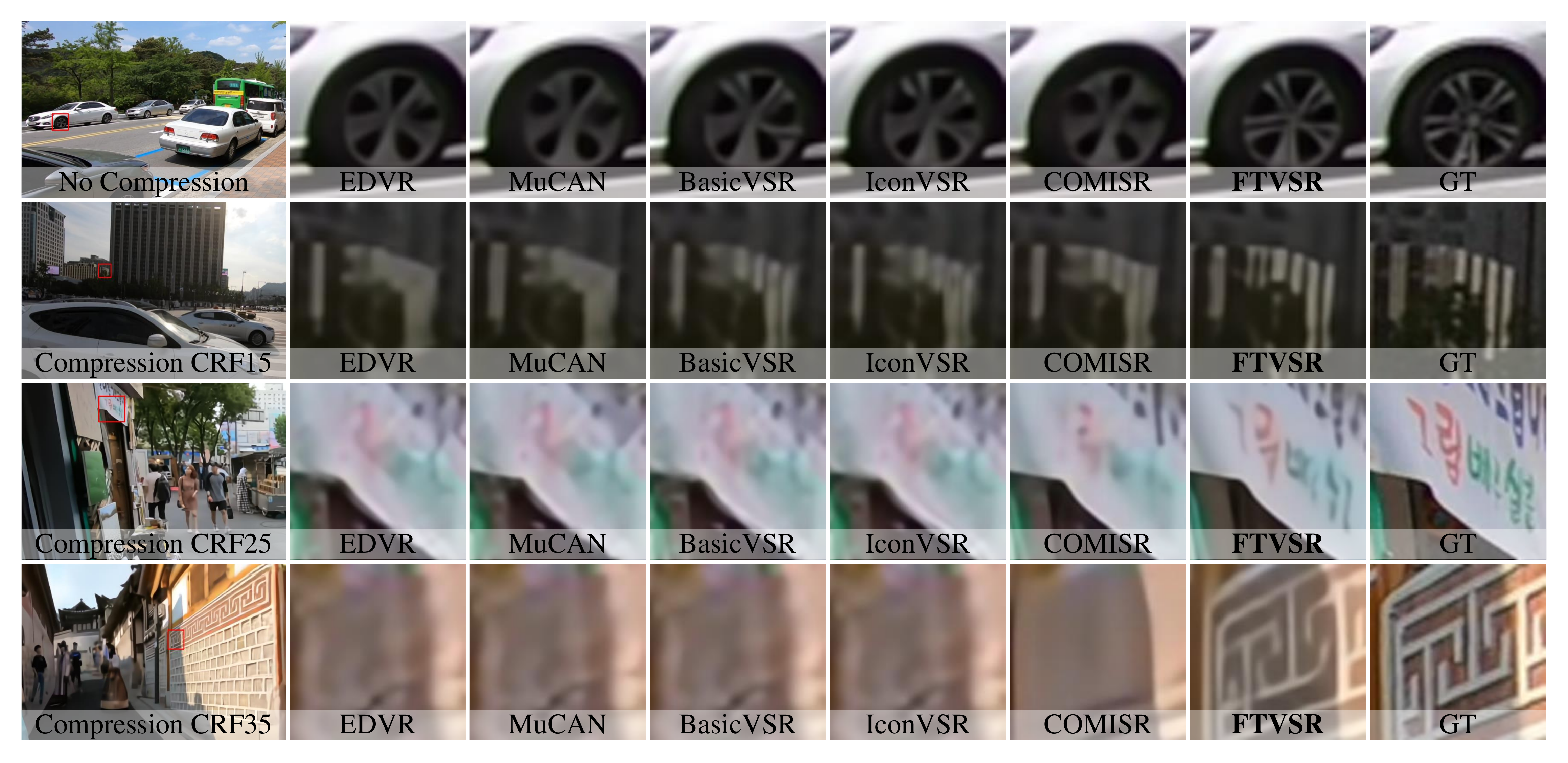}
% \vspace{-0.8cm}

\caption{The Visualization and comparison between FTVSR and other state-of-the-art VSR methods~\cite{wang2019edvr,li2020mucan,chan2021basicvsr,li2021comisr} on the uncompressed videos and compressed videos with compression rates of CRF 15, 25, and 35.}
\label{fig:case}
\end{figure*}

\subsubsection{Evaluation Metrics}
Following previous works~\cite{chan2021basicvsr,cao2021video,wang2019edvr,li2020mucan}, peak signal-to-noise ratio (PSNR) and structural similarity index (SSIM)~\cite{wang2004image} are used for evaluation.
To verify the effectiveness of FTVSR on videos with complicated degradation, we evaluate FTVSR on different degradation settings, including uncompressed VSR, compressed VSR, Blur-based VSR, Noise-based VSR, and Real-world VSR.

\textbf{Uncompressed VSR Evaluation:}
For the uncompressed videos, following the previous works~\cite{chan2021basicvsr,cao2021video,wang2019edvr,li2020mucan}, we train and test FTVSR on the low-resolution videos generated by bicubic downsampling from the videos in REDS and Vimeo-90K datasets.

\textbf{Compressed VSR Evaluation:} 
For compressed videos, we evaluate FTVSR on the compressed videos generated by the most common setting of H.264 codec with different compression algorithms. 
Following COMISR~\cite{li2021comisr}, we use the compression algorithm with CRF (Constant Rate Factor) of 15, 25, and 35 to generate compressed videos. We then evaluate FTVSR on these compressed videos and report the PSNR and SSIM. Besides, to evaluate the generalization ability of FTVSR on different compression algorithms, we also use other typical compression algorithms (CQP: Constant Quantization Parameter and CBR: Constant Bitrate) to generate compressed videos for evaluation.

\textbf{Blur-based VSR Evaluation:}
For the low-quality videos generated by downsampling with blur kernel, following~\cite{chan2021basicvsr}, we used the blur-based videos on REDS~\cite{nah2019ntire} dataset for training and evaluation.

\textbf{Noise-based VSR Evaluation:}
For the low-quality videos with additive noise, following~\cite{yang2022degradation}, we add Gaussian noise with noise level 15 on the videos generated by bicubic downsampling in REDS~\cite{nah2019ntire} for training and evaluation.

\textbf{Real-world VSR Evaluation:}
For the real-world VSR evaluation, following \cite{chan2022investigating}, we test FTVSR on the real-world VSR dataset: VideoLQ~\cite{chan2022investigating} and compare it with other VSR approaches~\cite{chan2021basicvsr,chan2022basicvsr++,chan2022investigating} by visualization since no ground truth on the real-world VSR dataset.

\subsection{Comparison with State-of-the-art Methods}

\subsubsection{Evaluation on Uncompressed Videos}
For the evaluation of uncompressed videos, REDS and Vid4 datasets are used for testing FTVSR, respectively. 
For a fair comparison with previous works~\cite{chan2021basicvsr,liu2022learning}, we compare with SOTA methods following the setting of training and testing on uncompressed videos generated by $4\times$ downsampling. Results are shown in Table \ref{tab_reds_unc}.
% , the results of other methods are cited from their papers.
FTVSR outperforms TTVSR~\cite{liu2022learning} on the REDS4 dataset and the Vid4 dataset. FTVSR achieves a gain of 0.3dB in PSNR on the REDS dataset and a gain of 0.3dB in PSNR on the Vid4 dataset, respectively. 
The SOTA VSR methods~\cite{chan2021basicvsr,liu2022learning} are designed for uncompressed videos, which usually performs not well on compressed videos. 
While COMISR~\cite{li2021comisr} is specifically designed for compressed videos, which usually performs well on compressed videos (introduced in section \ref{sec_eval_compress}), but not well on uncompressed videos. Compared with COMISR~\cite{li2021comisr}, FTVSR performs better on REDS and Vid4 datasets with gains of 2.74dB and 1.39dB and achieves SOTA results on uncompressed and compressed videos. The visualization comparisons on uncompressed videos with other methods are shown in Figure~\ref{fig:case}. FTVSR shows better visualization quality on uncompressed videos.

\subsubsection{Evaluation on Compressed Videos}
\label{sec_eval_compress}
To evaluate FTVSR on the compressed videos, we test FTVSR and make the comparison with other SOTA approaches on the REDS~\cite{nah2019ntire} dataset and Vid4~\cite{xue2019video} dataset, respectively. 
Same as the settings of compression in COMISR~\cite{li2021comisr}, compressed videos in several compression rates (CRF15, CRF25, CRF35) with the same compression algorithms (CRF) are generated. We evaluate FTVSR on these compressed videos and report PSNR and SSIM. 
To evaluate the generalization ability of FTVSR on other compression algorithms, we also compare FTVSR with other methods on the compressed videos generated by CQP and CBR compression modes.

\textbf{Different Compression Rates:}
The evaluation results of FTVSR on compressed videos from REDS~\cite{nah2019ntire} dataset are shown in Table \ref{tab_compress_reds}. For the compression results, we cite the number of other approaches from \cite{li2021comisr}. 
For a fair comparison, IconVSR and BasicVSR are finetuned with the same compression training process in \cite{li2021comisr}. 
Despite the remarkable results achieved on uncompressed videos, BasicVSR obtains 25.93dB and 23.22dB in PSNR on compressed videos of CRF25 and CRF35, respectively.
IconVSR only reaches 25.93dB and 23.22dB in PSNR on compressed videos of CRF25 and CRF35, which is the same as BasicVSR, despite outperforming it on uncompressed videos, where IconVSR performs better than BasicVSR.
These results show that the compression problems can not be solved by increasing the model capacity.

To alleviates the compression problem, COMISR~\cite{li2021comisr} designs a special compression-aware module, but it brings limited gains of 26.47 and 23.56dB in PSNR on the compressed videos of CRF25 and CRF35, respectively. 
However, our FTVSR obtains PSNR values of 30.55, 28.13dB, and 24.87dB on compressed videos of CRF15, CRF25, and CRF35, respectively. The fact that FTVSR outperforms COMISR by 1.6 dB on compressed videos of CRF25 indicates that our FTVSR is more capable of dealing with the issue of video compression.

The results on compressed videos of Vid4~\cite{xue2019video} dataset are shown in Table \ref{tab_com_vid4}. Similar to the settings on the REDS dataset, we finetune BasicVSR and IconVSR on the compressed videos of the Vimeo-90K dataset as \cite{li2021comisr}. 
On compressed videos of CRF 15, 25, and 35, FTVSR achieves 27.40dB, 25.38dB, and 22.61dB in PSNR, respectively. Additionally, FTVSR performs better than competing approaches and achieves new SOTA results, highlighting its enormous potential for overcoming the video compression dilemma.

The visualization results of FTVSR and other approaches on compressed videos are shown in Figure \ref{fig:case}. FTVSR shows better visual quality on compressed videos. Especially on the compressed video of CRF25 and CRF35, FTVSR has superior visual quality than other methods. 
Compared with the evaluations on compressed videos and uncompressed videos, COMISR~\cite{li2021comisr} performs better than BasicVSR~\cite{chan2021basicvsr} and IconVSR~\cite{chan2021basicvsr} on compressed videos, while performs poorly on uncompressed videos. In contrast, as shown in Figure \ref{fig:case}, FTVSR not only performs better on compressed videos but also performs well on uncompressed videos. Typically, BasicVSR, IconVSR, and COMISR fail to recover the texture on the compressed cases of CRF35, while our FTVSR still shows better visual quality in these cases. It profits from the frequency attention in FTVSR, which enables the generation of high-frequency textures with the guidance of low-frequency information.

\begin{table}[!t]
  \caption{Comparison with other methods on compressed videos of REDS dataset generated by different compression algorithms.}
%   \vspace{-0.2cm}
\tiny
  \centering
  \renewcommand\arraystretch{1.2}
  \resizebox{\columnwidth}{!}{
  \begin{tabular}{ l|ccc}
  \hline
  
  \hline
  Methods & CRF 25 & CQP 25 & CBR 1500k\\
    \hline
    RealBasicVSR~\cite{chan2022investigating} & 26.57/0.760 & 25.42/0.710 & 26.90/0.774\\
  \hline
  EDVR~\cite{wang2019edvr} & 25.98/0.706 & 26.46/0.718 & 29.55/0.836\\
  BasicVSR~\cite{chan2021basicvsr} & 25.93/0.704 & 26.62/0.722 & 30.41/0.858\\
  IconVSR~\cite{chan2021basicvsr} & 25.93/0.704 & 26.64/0.772 & 30.52/0.860\\
  BasicVSR++~\cite{chan2022basicvsr++} & 25.91/0.702 & 26.67/0.723 & 31.01/0.860\\
  TTVSR~\cite{liu2022learning} & 25.98/0.706 & 26.65/0.723 & 30.80/0.864\\

  \hline
  FTVSR & \textBF{28.13/0.778} & \textBF{28.77/0.793} & \textBF{31.08/0.865}\\
    \hline 
    
    \hline
  \end{tabular}
  }
  \label{tab_compress_alg}
\end{table}

\begin{figure}[]
\centering
\includegraphics[width=\columnwidth]{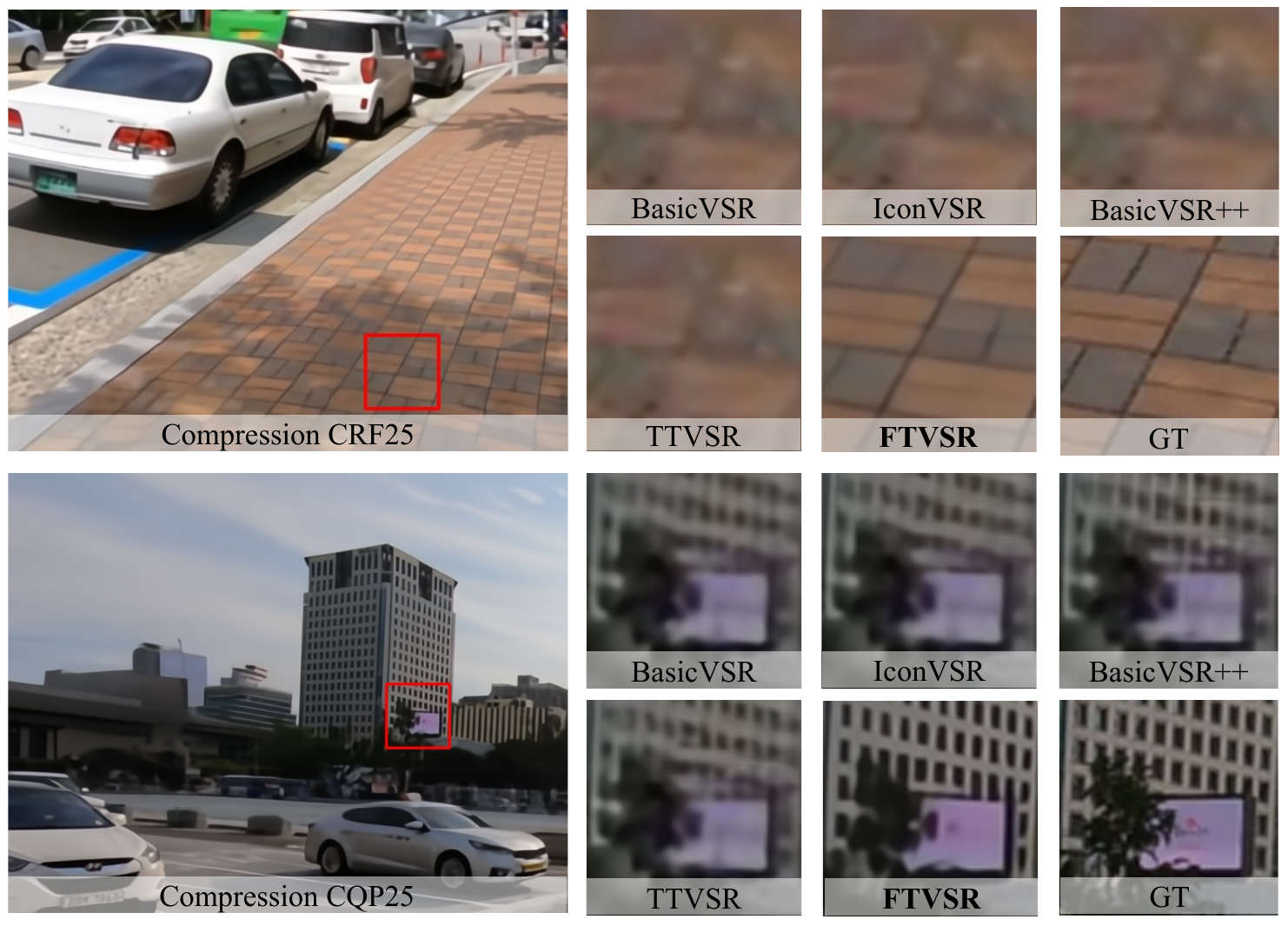}
\caption{Visualization results of BasicVSR~\cite{chan2021basicvsr}, IconVSR~\cite{chan2021basicvsr}, BasicVSR++~\cite{chan2022basicvsr++}, TTVSR~\cite{liu2022learning}, and our FTVSR on compressed videos generated by different compression algorithms.}
\label{fig:vis_case_crf_cqp}
\end{figure}

\textbf{Different Compression Algorithms:}
We compare FTVSR with other methods~\cite{wang2019edvr,chan2021basicvsr,chan2022basicvsr++,liu2022learning,chan2022investigating} on compressed videos generated by three different compression algorithms (CRF25, CQP25, and CBR1500k). The results are shown in Table \ref{tab_compress_alg}. 
In Table \ref{tab_compress_alg}, the results of other methods are obtained by finetuning their models on the compressed videos of REDS dataset. 
For the widely-used VSR methods (EDVR~\cite{wang2019edvr}, BasicVSR~\cite{chan2021basicvsr}, IconVSR~\cite{chan2021basicvsr}, BasicVSR++~\cite{chan2022basicvsr++}, and TTVSR~\cite{liu2022learning}), which have no designs for compression problem and perform worse on the compressed videos.
RealBasicVSR~\cite{chan2022investigating} is designed for the real-world VSR, which adds a cleaning module to remove artifacts before applying the VSR model. But RealBasicVSR just outperforms other VSR approaches on compressed videos in CRF25, while remains the gaps of 1.2dB and 3.1dB in PSNR on the compressed videos in CQP25 and CBR1500k, respectively. This is because the clean module cannot remove the artifacts caused by compression. Moreover, This phenomenon shows that the pipeline of cleaning first and applying the VSR model later has the limitation of generalization. However, FTVSR achieves 28.05, 28.77, and 31.08dB in PSNR on the compressed videos with three different compression algorithms, respectively. These results show that FTVSR has a superior ability on handling compression problems in VSR.

The visualization of compressed videos generated by different compression algorithms is shown in Figure \ref{fig:vis_case_crf_cqp}. Compared with other VSR methods~\cite{chan2021basicvsr,chan2022basicvsr++,liu2022learning}, FTVSR shows better visualization results since the frequency attention mechanism.

\begin{table}[!t]
  \caption{Comparison with other methods on the videos of REDS dataset with complicated degradations. 'BI' means bicubic downsampling. 'BD' means downsampling with a blur kernel. 'RG-Noise' means adding random Gaussian noise.}
%   \vspace{-0.2cm}
% \tiny
  \centering
  \renewcommand\arraystretch{1.2}
    \renewcommand\tabcolsep{9pt}
  \resizebox{\columnwidth}{!}{
  \begin{tabular}{ l|ccc}
  \hline
  
  \hline
  Methods & BI & BD & RG-Noise\\
    \hline
  RealBasicVSR~\cite{chan2022investigating} & 27.04/0.780 & 26.70/0.783 & 26.07/0.736\\
  \hline
  EDVR~\cite{wang2019edvr}& 31.09/0.880 & 34.73/0.925 & 28.74/0.810\\
  BasicVSR~\cite{chan2021basicvsr} & 31.42/0.890 & 35.23/0.931 & 29.60/0.841\\
  IconVSR~\cite{chan2021basicvsr} & 31.67/0.895 & 35.37/0.933 & 29.88/0.848\\
  TTVSR~\cite{liu2022learning} & 32.12/0.901 & 35.54/0.935 & 29.81/0.846\\
  BasicVSR++\cite{chan2022basicvsr++}& 32.38/0.906 & 35.57/0.935 & 30.20/0.854 \\

  \hline
  FTVSR & \textBF{32.42/0.907} & \textBF{37.81/0.949} & \textBF{30.48/0.852}\\
    \hline 
    
    \hline
  \end{tabular}
  }
  \label{tab_bd_noise}
\end{table}

\begin{figure}[!t]
\centering
\includegraphics[width=\columnwidth]{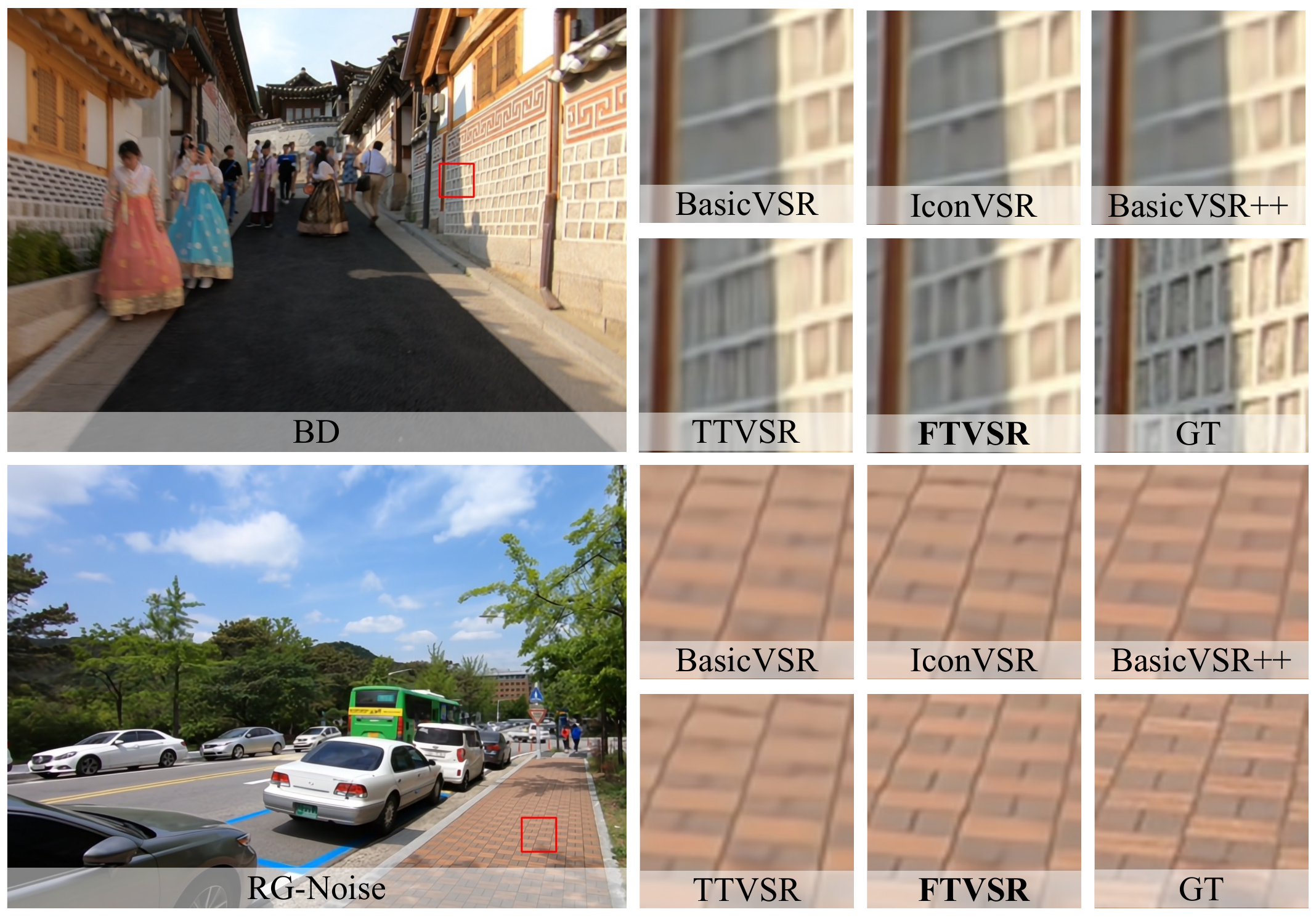}
\caption{Visualization results of BasicVSR~\cite{chan2021basicvsr}, IconVSR~\cite{chan2021basicvsr}, BasicVSR++~\cite{chan2022basicvsr++}, TTVSR~\cite{liu2022learning}, and our FTVSR on the LR videos with different degradations (BD and random Gaussian noises with a scale parameter of 15).}
\label{fig:vis_cases_bd_noise}
\end{figure}

\begin{figure}[!t]
\centering
\includegraphics[width=\columnwidth]{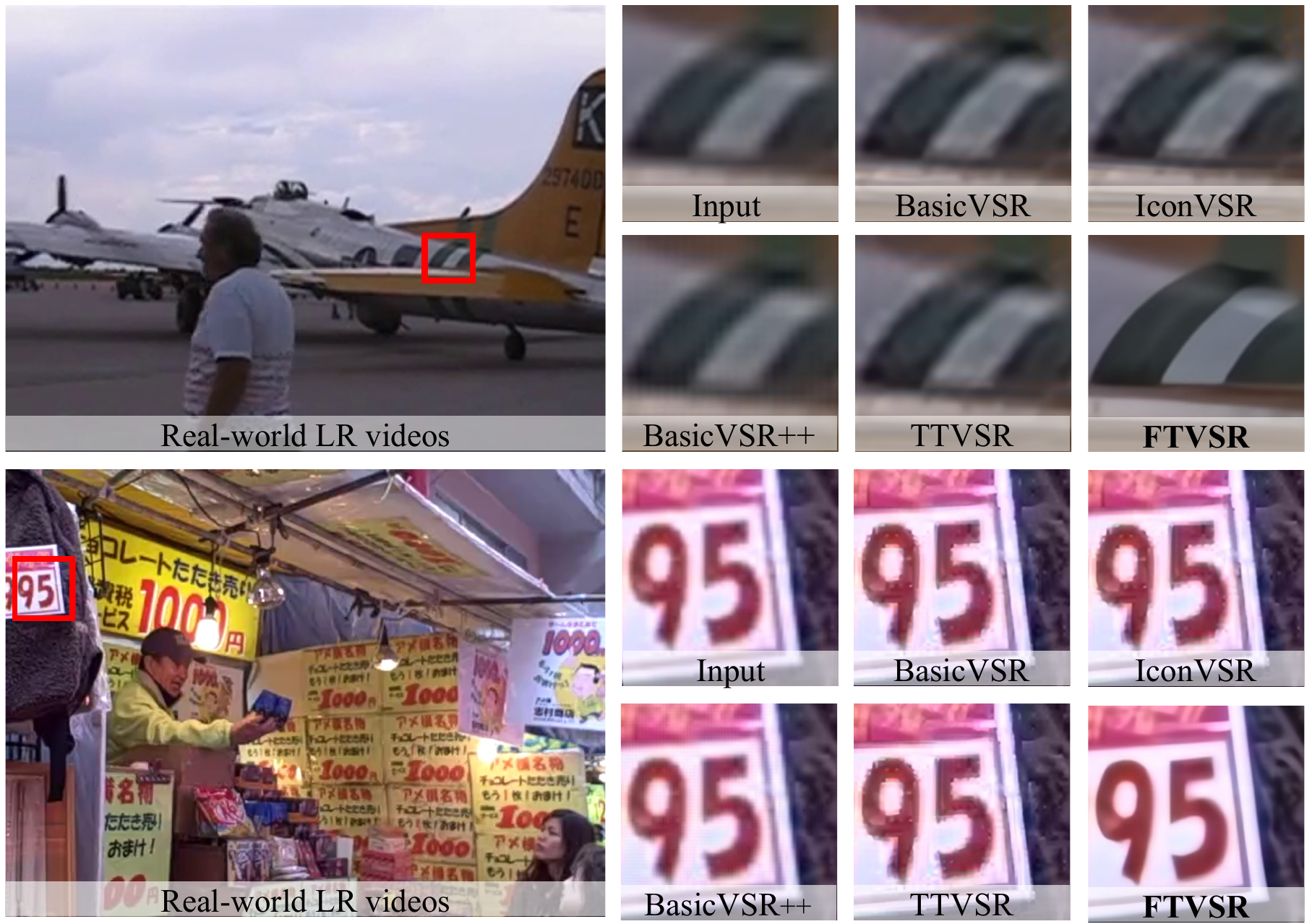}
\caption{Visualization results of BasicVSR~\cite{chan2021basicvsr}, IconVSR~\cite{chan2021basicvsr}, BasicVSR++~\cite{chan2022basicvsr++}, TTVSR~\cite{liu2022learning}, and our FTVSR on the real-world LR videos from VideoLQ dataset~\cite{chan2022investigating}.}
\label{fig:real_world}
\end{figure}

\subsubsection{Evaluation on Videos with Blur}
\label{sec_blur}
To verify the ability of FTVSR on handling the complicated degradation with blur kernel, we compare FTVSR with EDVR~\cite{wang2019edvr}, BasicVSR~\cite{chan2021basicvsr}, IconVSR~\cite{chan2021basicvsr}, BasicVSR++~\cite{chan2022basicvsr++}, TTVSR~\cite{liu2022learning}, and RealBasicVSR~\cite{chan2022investigating} in Table \ref{tab_bd_noise}. 
For the different downsampling kernels (BI and BD), FTVSR achieves 32.42dB and 37.81dB in PSNR, which outperforms other methods and achieves SOTA results. These results show that FTVSR can handle complicated degradations. 
The results of RealBasicVSR~\cite{chan2022investigating} show that the clean mechanism in RealBasicVSR cannot handle these complicated degradations with different kernels.
Besides, the visualization results on LR videos with BI and BD kernels are shown in Figure \ref{fig:vis_cases_bd_noise}. Compared with other methods, FTVSR achieves superior visualization results.

\subsubsection{Evaluation on Videos with Noise}
\label{sec_noise}
To verify the ability of FTVSR to handle the complicated degradation with additive noise, we compare FTVSR with EDVR~\cite{wang2019edvr}, BasicVSR~\cite{chan2021basicvsr}, IconVSR~\cite{chan2021basicvsr}, BasicVSR++~\cite{chan2022basicvsr++}, TTVSR~\cite{liu2022learning}, and RealBasicVSR~\cite{chan2022investigating} in Table \ref{tab_bd_noise}. For a fair comparison, we add random Gaussian noises with a scale parameter of 15 on LR videos. As shown in Table \ref{tab_bd_noise}, FTVSR achieves SOTA results with PSNR of 30.48dB in the setting of RG-Noise. Although RealBasicVSR~\cite{chan2022investigating} uses a clean module to remove the noise, it only achieves 26.07dB in PSNR. Therefore, FTVSR has better generalization ability compared with RealBasicVSR.
Besides, the visualization results on LR videos with noises are shown in Figure \ref{fig:vis_cases_bd_noise}. Compared with other methods, FTVSR achieves superior visualization results.

\begin{table}[]
  \caption{The comparison of parameters and FLOPs.}
%   \vspace{-0.2cm}
% \tiny
  \centering
  \renewcommand\arraystretch{1.2}
    \renewcommand\tabcolsep{11pt}
  \resizebox{\columnwidth}{!}{
  \begin{tabular}{ l|ccc}
  \hline
  
  \hline
  Methods & Params(M) & FLOPs(T) & PSNR/SSIM\\
    \hline
DUF~\cite{jo2018deep} & 5.8 & 2.34 & 24.19/0.692\\
EDVR~\cite{wang2019edvr} & 20.6 & 2.95 & 25.98/0.706\\
MuCAN~\cite{li2020mucan} & 13.6 & $>$1.07 & 25.96/0.705\\
BasicVSR~\cite{chan2021basicvsr} & 6.3 & 0.33 & 25.93/0.704\\
IconVSR~\cite{chan2021basicvsr} & 8.7 & 0.51 & 25.93/0.704\\
TTVSR~\cite{liu2022learning} & 6.8 & 0.61 & 25.98/0.706\\
COMISR~\cite{li2021comisr} & 6.2 & 0.36 & 26.47/0.728\\
  \hline
  FTVSR & 10.8 & 0.76 & 27.28/0.763\\
    \hline 
    
    \hline
  \end{tabular}
  }
  \label{tab_params}
\end{table}

\begin{figure*}[]
\centering
\includegraphics[width=0.9\textwidth]{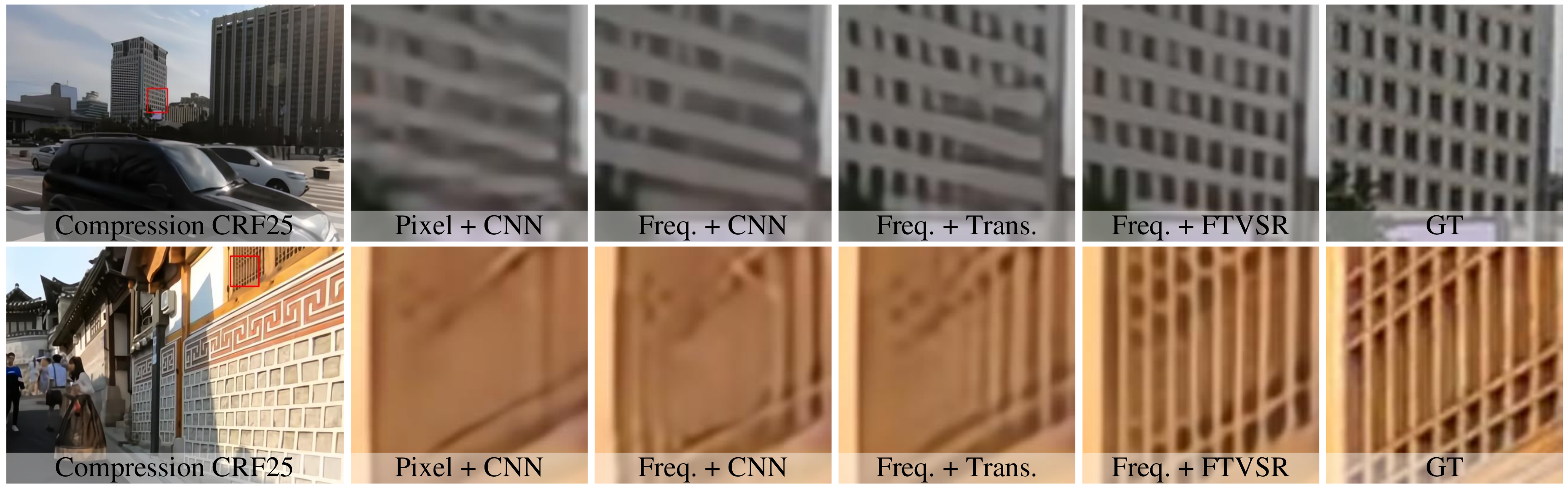}
\caption{Visualization results of ``Pixel + CNN", ``Frequency + CNN", ``Frequency + Transformer", and ``Frequency + FTVSR" in Table \ref{tab_freq}. The frequency learning mechanism makes the texture clear.}
\label{fig:ablation}
\end{figure*}

\begin{table*}[!t]
  \caption{The ablation study of learning in the frequency domain. Each entry shows PSNR$\uparrow$/SSIM$\uparrow$ on the REDS4 dataset.}
%   \vspace{-0.2cm}
  \centering
  \tiny
    \renewcommand\arraystretch{1.2}
  \resizebox{\textwidth}{!}{
  \begin{tabular}{ l | c | c | c | c | c | c | c}
    \hline
    
    \hline
    \multirow{2}{*}{Domain + Backbone} & \multicolumn{3}{c|}{Average of clips with Compression} & \multicolumn{4}{c}{Per clip with Compression CRF25} \\
    \cline{2-8}
     & CRF15 & CRF25 & CRF35 & Clip\_000 &  Clip\_011 &  Clip\_015 &  Clip\_020  \\
    \hline
    Pixel + CNN & 29.05/0.814 & 25.93/0.704 & 23.22/0.596 & 24.37/0.628 & 26.01/0.702 & 28.13/0.777 & 25.21/0.709 \\
    Frequency + CNN & 29.20/0.825 & 26.87/0.745 & 23.83/0.629 & 24.98/0.666& 27.11/0.746 & 29.36/0.818 & 26.05/0.751  \\
    
    Frequency + Transformer & 29.51/0.837 & 27.15/0.759 & 24.03/0.644 & 25.20/0.684 & 27.53/0.763 & 29.47/0.828 & 26.33/0.766  \\
    Frequency + FTVSR & \textBF{29.70/0.843} & \textBF{27.28/0.763} & \textBF{24.22/0.646} & \textBF{25.26/0.609} & \textBF{27.75/0.766} & \textBF{29.62/0.831} & \textBF{26.47/0.772} \\
    \hline
    
    \hline
  \end{tabular}
  }
  \label{tab_freq}
%   \vspace{-0.3cm}
\end{table*}

\subsubsection{Evaluation on Real-world Scenes}
\label{sec_real_world}
To verify the generalization ability of FTVSR on real-world videos, we evaluate FTVSR and compare it with other methods (BasicVSR~\cite{chan2021basicvsr}, IconVSR~\cite{chan2021basicvsr}, BasicVSR++~\cite{chan2022basicvsr++}, TTVSR~\cite{liu2022learning}, and RealBasicVSR~\cite{chan2022investigating}) on VideoLQ~\cite{chan2022investigating} dataset. VideoLQ dataset is a real-world VSR dataset proposed by RealBasicVSR~\cite{chan2022investigating}, which contains low-quality videos from websites and youtube. For a fair comparison, we use FTVSR trained on the REDS dataset and all other models are trained on the REDS dataset. The visualization results are shown in Figure \ref{fig:real_world} since no ground truth in the real-world VSR dataset. FTVSR performs better than other methods and has a superior visual quality, even compared with the real-world VSR method RealBasicVSR~\cite{chan2022investigating}.

\subsubsection{Parameters and FLOPs}
The comparisons of parameters and FLOPs between FTVSR with other methods are shown in Table \ref{tab_params}. We compute the FLOPs based on the LR input frames of size $180\times 320$ and perform $4\times$ upsampling VSR. In Table \ref{tab_params}, FTVSR is based on the backbone of BasicVSR. FTVSR achieves 27.28/0.763 in PSNR/SSIM with 10.8M parameters and 0.76G FLOPs, which outperforms other SOTA methods with comparable parameters and FLOPs. Due to the FTVSR process HR frames in the frequency domain and DCT operation reducing the computational costs, the FLOPs of FTVSR are comparable with BasicVSR.

\subsection{Ablation Study}
In this section, we first introduce the comparison of Transformer and CNN in frequency learning in Section \ref{sec:ab_1}. Then, the ablation study of local, global, and dual frequency attention are shown in Section \ref{sec:ab_2}. Finally, we explore the different frequency attention mechanisms in a joint space-time-frequency domain for video processing in Section \ref{sec:ab_3}.

\subsubsection{Learning in Frequency Domain}
\label{sec:ab_1}
To verify the effectiveness of the insight that learning in the frequency domain, we perform the ablation study of learning in the frequency domain and compare CNN-based and Transformer-based methods in both pixel and frequency domains on the REDS4 dataset. The results are shown in Table \ref{tab_freq}. For the setting of Pixel+CNN, BasicVSR~\cite{chan2021basicvsr}, learning to restore HR frames in the pixel domain. On the compressed videos of CRF15, CRF25, and CRF35, this model obtains PSNR of 29.05, 25.93, and 23.22dB, respectively.
Compared with its 31.42dB obtained on uncompressed videos, it shows the poor capacity of leaning in the pixel domain by CNN-based network. 
After transferring images into the frequency domain, on the compressed movies of CRF25, the Frequency+CNN setting yields a relative boost in PSNR of 0.94dB.
Moreover, in the frequency domain, combined with a simple Transformer-based model without frequency attention, the setting of Frequency+Transformer achieves 27.15dB in PSNR on the compressed videos of CRF25.
Compared with the pixel domain and frequency domain, learning in the frequency domain shows better results.
Compared with CNN and Transformer, the Transformer in the frequency domain shows better results.
Replacing the basic attention mechanism in the Transformer by frequency attention in FTVSR, the setting of Frequency+FTVSR achieves 27.28dB in PSNR on the compressed videos of CRF25, which shows that the frequency attention in FTVSR has a stronger capacity than vanilla attention in the frequency domain. 
Compared to the visualization results in Figure \ref{fig:ablation}, the setting of Frequency+FTVSR demonstrates superior visual quality than others.

\begin{table}[!t]
  \caption{Ablation study of LFA, GFA, and DFA on the low-quality LR videos from REDS dataset with different degradations. Note that FTVSR in the conference is based on LFA.}
%   \vspace{-0.2cm}
% \tiny
  \centering
  \renewcommand\arraystretch{1.2}
    \renewcommand\tabcolsep{9pt}
  \resizebox{\columnwidth}{!}{
  \begin{tabular}{ c|ccc}
  \hline
  
  \hline
  Frequency Attention & LFA~\cite{qiu2022learning} & GFA & DFA\\
    \hline 
  CRF25 & 27.23/0.761 & 27.14/0.759 & \textBF{27.53/0.765} \\
  BD & 35.82/0.930 & 35.87/0.936 & \textBF{36.33/0.939} \\
  RG-Noise & 26.05/0.677 & 26.09/0.680 & \textBF{26.35/0.685} \\

    \hline 
    
    \hline
  \end{tabular}
  }
  \label{tab_ab_dfa}
\end{table}

\begin{figure}[]
\centering
\includegraphics[width=\columnwidth]{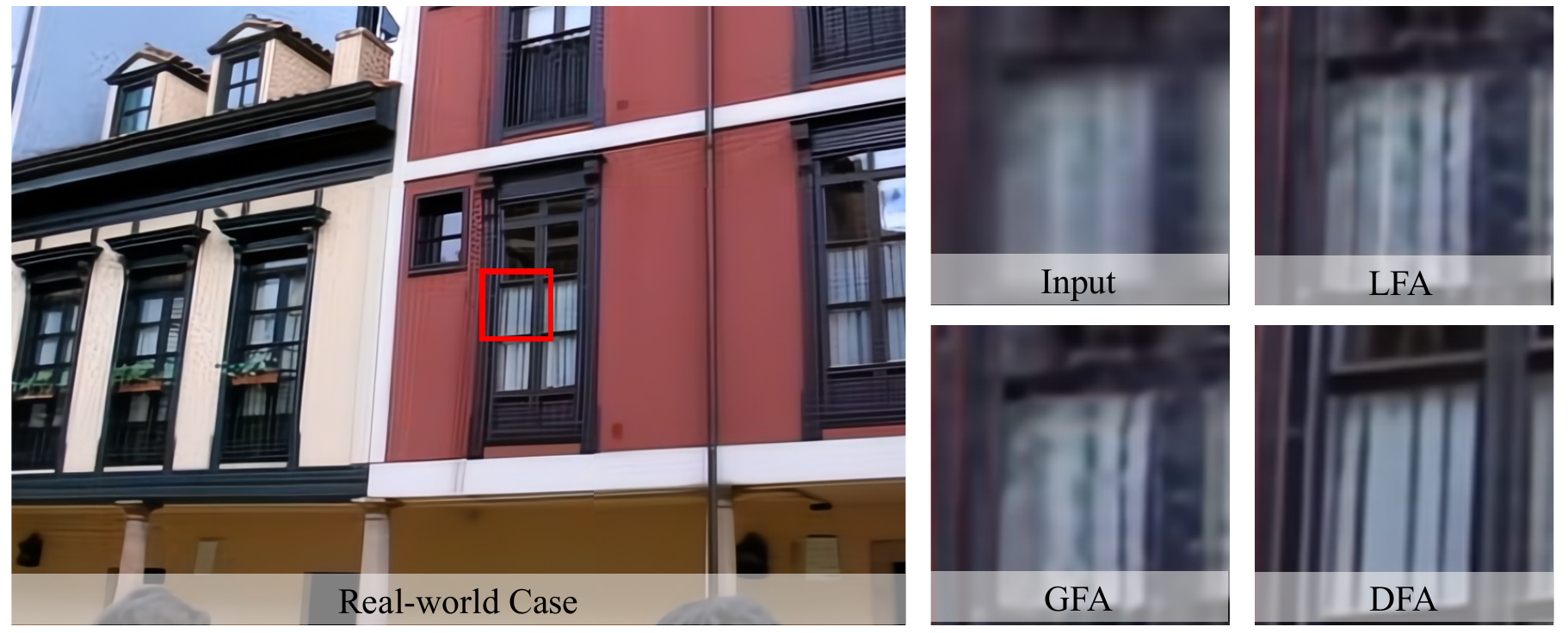}
\caption{Visualization results of LFA, GFA, and DFA on the real-world case from \cite{chan2022investigating}.}
\label{fig:comare_dfa}
\end{figure}

\subsubsection{Local and Global Frequency Relation}
\label{sec:ab_2}
To evaluate the effectiveness of capturing local and global frequency relations, we compare local frequency attention (LFA), global frequency attention (GFA), and dual frequency attention (DFA) in Table \ref{tab_ab_dfa}. 
For the ablation study in Table \ref{tab_ab_dfa}, FTVSR is trained on compressed and uncompressed videos and evaluated on low-quality LR videos with various degradations (CRF25, BD, and RG-Noise). For the comparison of LFA and GFA, LFA performs better on compressed videos, while GFA performs better on the BD and RG-Noise videos. Moreover, DFA achieves the best results by the PSNR of 27.63dB, 36.33dB, and 26.35dB on CRF25, BD, and RG-Noise, respectively. These results demonstrate that the global and local frequency relations are essential for low-quality VSR. 

Compared with our previous work~\cite{qiu2022learning} in the conference that adopts LFA as the attention block in Transformer and achieves 27.23dB on CRF25 in the ablation study, the new DFA in FTVSR brings an improvement of 0.3dB and achieves 27.53dB. On the videos with blur and noise, the new FTVSR also performs better than the conference version. Besides, the visual comparison of LFA, GFA, and DFA in the real-world case is shown in Figure\ref{fig:comare_dfa}. DFA has better visualization results than LFA and GFA in handling complicated degradations in real-world scenarios.

\begin{table}[!t]
  \caption{The ablation study of different frequency attention mechanisms. ``Base" means vanilla attention without frequency attention.}
%   \vspace{-0.2cm}
\tiny
  \centering
  \renewcommand\arraystretch{1.3}
  \renewcommand\tabcolsep{8pt}
  \resizebox{\columnwidth}{!}{
  \begin{tabular}{ l|ccc}
    \hline

    \hline
    Type & CRF15 & CRF25 & CRF35\\
    \hline
    Base & 29.51/0.837 & 27.15/0.759 & 24.03/0.644 \\
    $\Lambda_S$ & 29.63/0.840 & 27.23/0.761 & 24.12/0.646 \\
    $\Lambda_T$ & 29.60/0.840 & 27.11/0.760 & 24.05/0.641\\
    $\Lambda_{T\times S}$& 29.61/0.839 & 27.22/0.760 & 24.11/0.644\\
    $\Lambda_{TS}$& 29.65/0.841 & 27.24/0.762 & 24.12/0.645\\
    $\Lambda_{ST}$& \textBF{29.70/0.843} & \textBF{27.28/0.763} & \textBF{24.22/0.646}\\
    %  Frequency Attention & Base & $\Lambda_S$ &  $\Lambda_T$ &  $\Lambda_{T\times S}$ &  $\Lambda_{TS}$ &  $\Lambda_{ST}$ \\
    %  \hline 
    %  Compression CRF15& 29.51/0.837 & 29.63/0.840 & 29.60/0.840 &29.61/0.839 & 29.65/0.841 & \textbf{29.70/0.843}\\
    
    %  Compression CRF25& 27.15/0.759 & 27.23/0.761 & 27.11/0.760 & 27.22/0.760& 27.24/0.762 &\textbf{27.28/0.763}  \\
    
    %  Compression CRF35& 24.03/0.644 & 24.12/0.646 & 24.05/0.641 &24.11/0.644 & 24.12/0.645 &\textbf{24.22/0.646} \\

    \hline
    
    \hline
  \end{tabular}
  }
  \label{tab_abattention}
%   \vspace{-0.3cm}
\end{table}

\subsubsection{Frequency Attention in Video}
\label{sec:ab_3}
We conduct the ablation study and make the comparison of different types of frequency attention on the REDS dataset to evaluate the effectiveness of various frequency attention on capturing spatial and temporal information in the frequency domain, which is introduced in Section \ref{frequecny_attention}.
% To evaluate the effectiveness of different frequency attention on capturing spatial and temporal information in the frequency domain, which is introduced in Section \ref{frequecny_attention}, we conduct the ablation study and make the comparison on the REDS dataset. 
The comparisons are shown in Table \ref{tab_abattention}. The basic Transformer with ``Base" attention mechanism computes spatial self-attention in the frequency domain, which is the baseline.
All FTVSR models in Table \ref{tab_abattention} are based on local frequency attention (LFA).
Compared with the different frequency attention mechanisms, including space-frequency attention $\Lambda_S$, time-frequency attention $\Lambda_T$, joint time-space-frequency attention $\Lambda_{T\times S}$, and divided time-space-frequency attention ($\Lambda_{TS}$ and $\Lambda_{ST}$), the ablation studies in Table \ref{tab_abattention} demonstrate that $\Lambda_{ST}$ achieves the best results on three types of compressed videos (CRF15, CRF25, and CRF35), respectively. 
The divided time-space-frequency $\Lambda_{ST}$ computes the space-frequency attention first and follows the time-frequency attention. This is because that degraded frames should be first recovered by the space-frequency attention ($\Lambda_{ST}$) and then restored by the time-frequency attention ($\Lambda_{TS}$). The recovered textures in $\Lambda_{ST}$ will benefit time-frequency attention for temporal learning.

% \setlength{\tabcolsep}{4pt}
% \begin{table}
% \begin{center}
% \caption{Font sizes of headings. Table captions should always be
% positioned {\it above} the tables. The final sentence of a table
% caption should end without a full stop}
% \label{table:headings}
% \begin{tabular}{lll}
% \hline\noalign{\smallskip}
% Heading level & Example & Font size and style\\
% \noalign{\smallskip}
% \hline
% \noalign{\smallskip}
% Title (centered)  & {\Large \bf Lecture Notes \dots} & 14 point, bold\\
% 1st-level heading & {\large \bf 1 Introduction} & 12 point, bold\\
% 2nd-level heading & {\bf 2.1 Printing Area} & 10 point, bold\\
% 3rd-level heading & {\bf Headings.} Text follows \dots & 10 point, bold
% \\
% 4th-level heading & {\it Remark.} Text follows \dots & 10 point,
% italic\\
% \hline
% \end{tabular}
% \end{center}
% \end{table}
% \setlength{\tabcolsep}{1.4pt}

\section{Conclusions}
In this study, we provide a unique spatiotemporal Frequency-Transformer (FTVSR) for low-quality Video Super-Resolution and reveal an insight of processing VSR in the frequency domain.
We convert video frames into the frequency domain and extract frequency tokens from the deep features to address the challenging degradation problem in the real world.
We introduce frequency attention, which captures frequency interactions across multiple frequency bands.
Frequency-based tokenization and frequency attention can benefit the generation of high-frequency textures with the guidance of low-frequency information.
Moreover, we propose dual frequency attention to capture the local and global frequency relations, which strengthens the capacity of FTVSR on handling complicated degradations.
To improve the visual quality of restored HR videos, we further explore frequency attention in a joint space-time-frequency domain. Extensive experiments and visualization results on several widely-used VSR datasets and real-world VSR benchmarks demonstrate the superior performance of the proposed FTVSR, which leads FTVSR to achieve state-of-the-art results.

\ifCLASSOPTIONcaptionsoff
  \newpage
\fi



%\bibliographystyle{IEEEtran}


% \end{thebibliography}

\bibliographystyle{IEEEtran}
\bibliography{ref}


\begin{IEEEbiography}[{\includegraphics[width=1in,height=1.25in,clip,keepaspectratio]{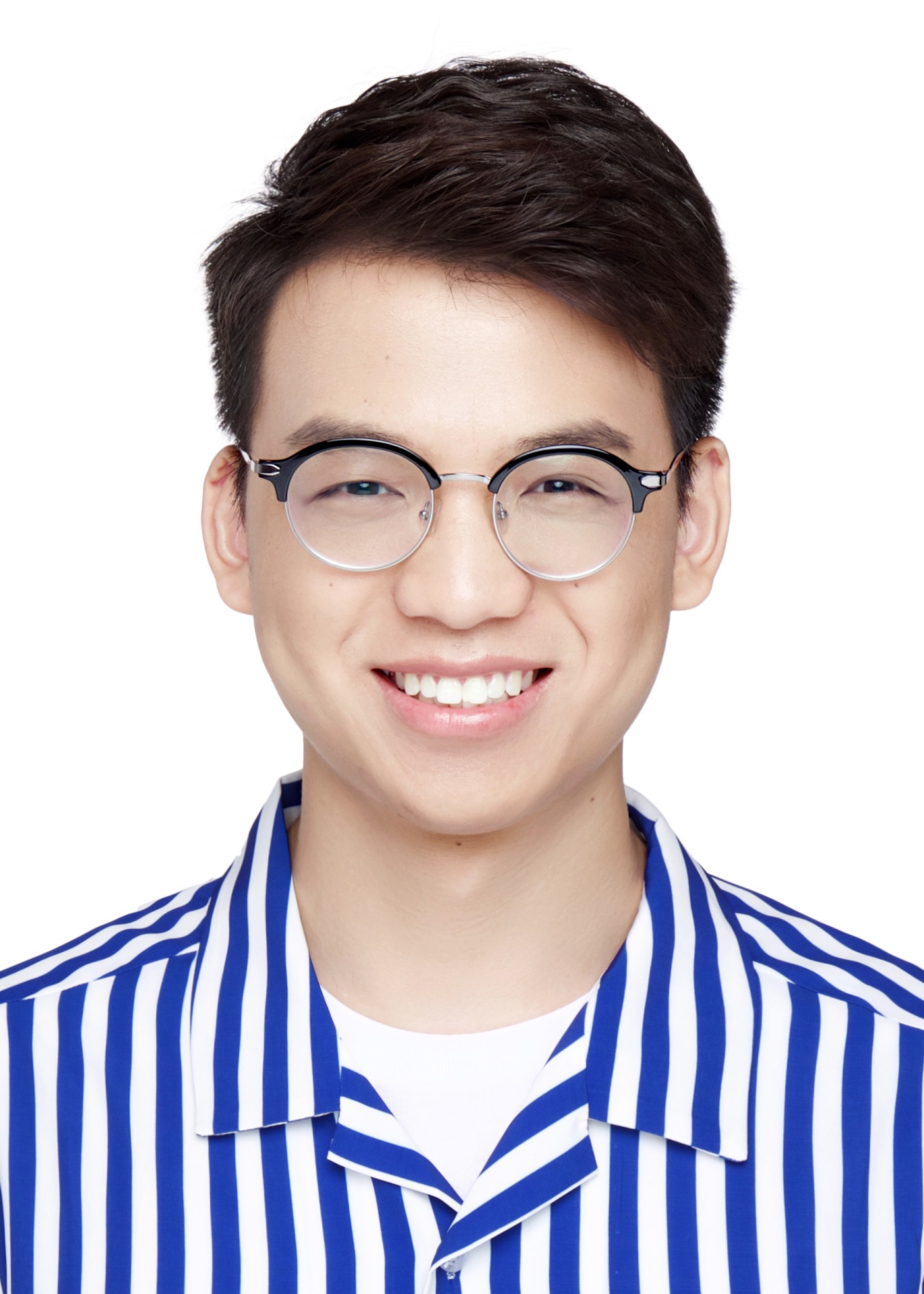}}]{Zhongwei Qiu}
Zhongwei Qiu received his B.S. degree in automation from the University of Science and Technology Beijing, China, in 2018, where he is currently pursuing the Ph.D. degree with the School of Automation and Electrical Engineering. He is also a visiting research student at the University of Sydney.
His research interests include 2D/3D human pose estimation and Video Super-Resolution.
\end{IEEEbiography}

\begin{IEEEbiography}[{\includegraphics[width=1in,height=1.25in,clip,keepaspectratio]{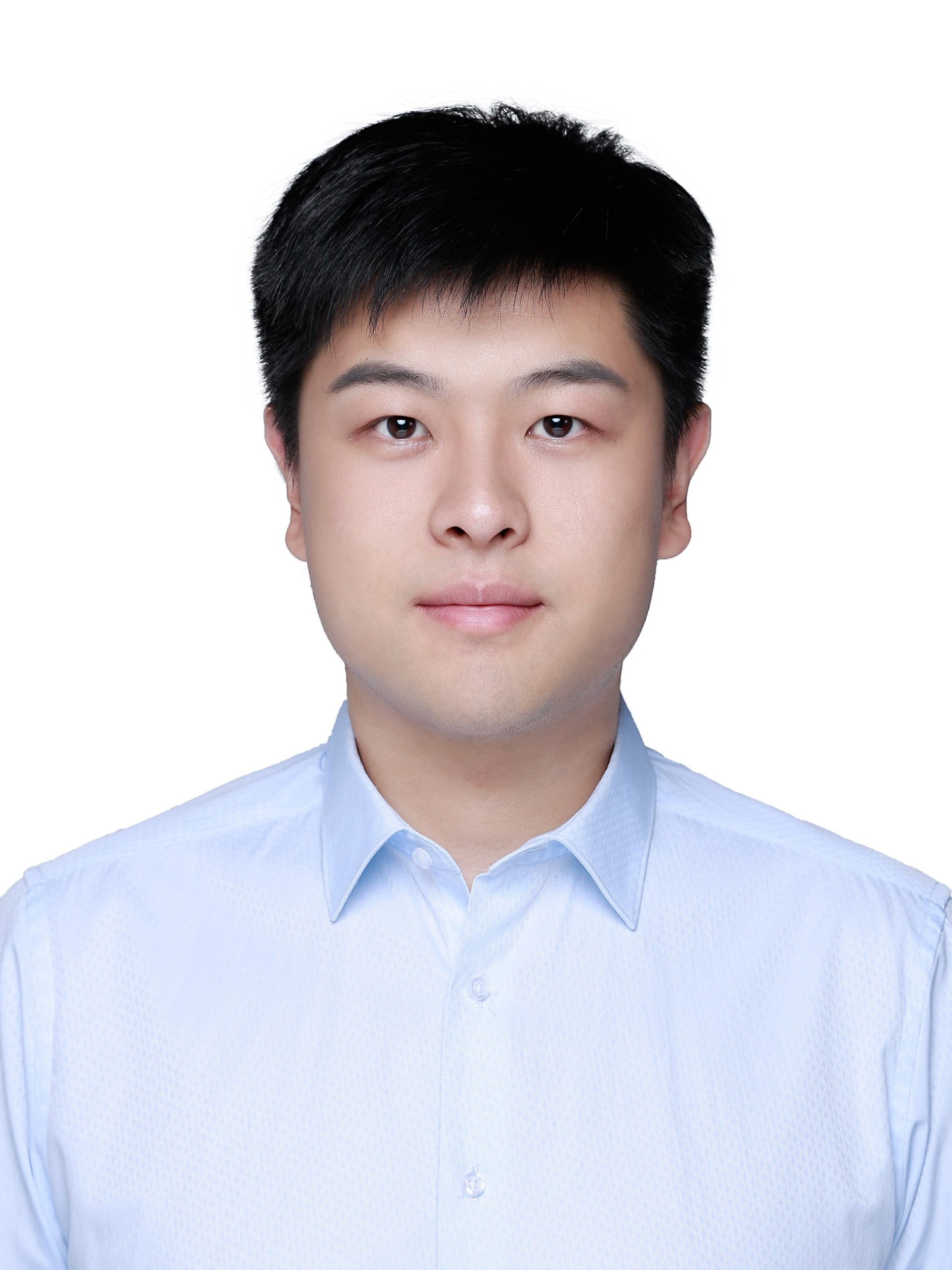}}]{Huan Yang}
Huan Yang received his BS and Ph.D. degrees in computer science in 2014 and 2019 respectively from Shanghai Jiao Tong University, China. He is currently a researcher at Microsoft Research Asia. His current research interest is image and video synthesis including enhancement, restoration, and generation.
\end{IEEEbiography}

\begin{IEEEbiography}[{\includegraphics[width=1in,height=1.25in,clip,keepaspectratio]{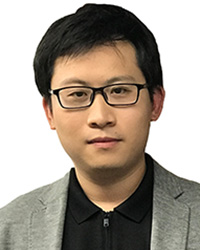}}]{Jianlong Fu}
  Jianlong Fu is currently a Lead Researcher with the Multimedia Search and Mining Group, Microsoft Research Asia. He received his Ph.D. degree in pattern recognition and intelligent system from the Institute of Automation, Chinese Academy of Science in 2015. His current research interests include computer vision, computational photography, vision and language. He has authored or coauthored more than 40 papers in journals and conferences, and 1 book chapter. He serves as a Lead organizer and guest editor of IEEE Trans. Pattern Analysis and Machine Intelligence Special Issue on Fine-grained Categorization. He is an area chair of ACM Multimedia 2018, ICME 2019. He received the Best Paper Award from ACM Multimedia 2018, and has shipped core technologies to a number of Microsoft products, including Windows, Office, Bing Multimedia Search, Azure Media Service and XiaoIce.
\end{IEEEbiography}

\begin{IEEEbiography}[{\includegraphics[width=1in,height=1.25in,clip,keepaspectratio]{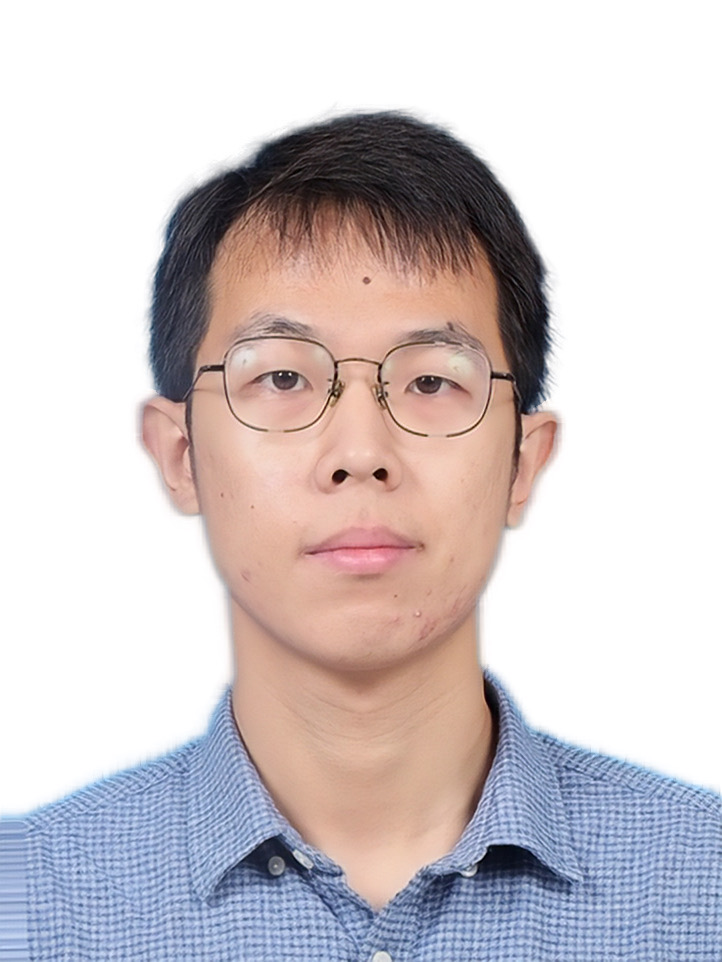}}]{Daochang Liu}
Daochang Liu is currently a postdoctoral researcher at the University of Sydney. He received a Ph.D. degree from Peking University in 2022, and a B.E. degree from Tongji University in 2017. His research interests include generative learning, video understanding, and surgical artificial intelligence.
\end{IEEEbiography}

\begin{IEEEbiography}[{\includegraphics[width=1in,height=1.25in,clip,keepaspectratio]{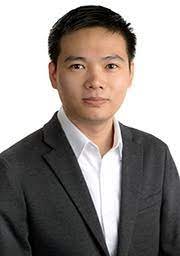}}]{Chang Xu}
Chang Xu received the PhD degree from Peking University, China. He is currently a senior lecturer and ARC DECRA fellow with the School of Computer Science, University of Sydney. He has authored or coauthored more than 100 papers in prestigious journals and top tier conferences. His research interests include machine learning algorithms and related applications in computer vision. He was the recipient of several paper awards, including Distinguished Paper Award in IJCAI 2018. He was the PC member or senior PC member for many conferences, including NeurIPS, ICML, ICLR, CVPR, ICCV, IJCAI, and AAAI. He has been recognized as Top Ten Distinguished Senior PC Member in IJCAI 2017.
\end{IEEEbiography}

\begin{IEEEbiography}[{\includegraphics[width=1in,height=1.25in,clip,keepaspectratio]{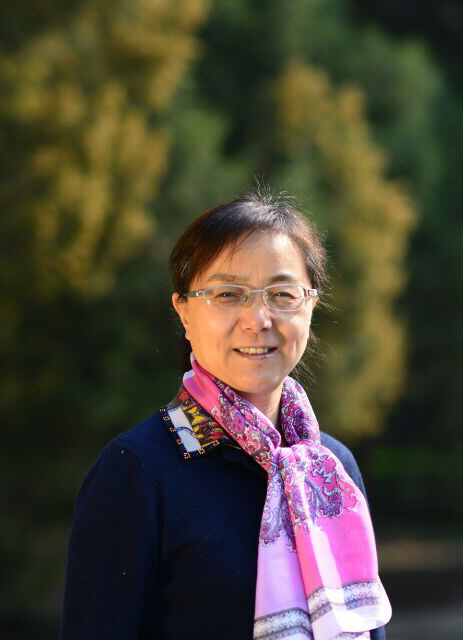}}]{Dongmei Fu}
Dongmei Fu received the M.S. degree from Northwestern Polytechnical University, in 1984, and the Ph.D. degree in automation science from the University of Science and Technology Beijing (USTB), China, in 2006, where she is currently a Professor and a Doctoral Supervisor. She has taken charge of several national projects about corrosion data mining and infrared image processing. Her current research interests include automation control theory, image processing, and data mining.
\end{IEEEbiography}


%\vfill


\end{document}